\documentclass[a4paper,11pt,unpublished]{quantumarticle}
\pdfoutput=1
\usepackage[utf8]{inputenc}
\usepackage[english]{babel}
\usepackage[T1]{fontenc}
\usepackage{amsmath}
\usepackage{hyperref}
\usepackage{multirow, booktabs}
\usepackage{float}
\usepackage{caption}
\usepackage{subcaption}
\usepackage{comment}
\usepackage{soul}
\usepackage{tikz}
\usepackage{lipsum}

\usepackage{adjustbox}
\usepackage{xcolor}
\usepackage{tabularray}
\definecolor{tableheader}{HTML}{FF6361}
\usepackage{graphicx}
\usepackage{textgreek}

\title{High PDMR contrast in single NV centres and related photocurrent properties}
\author[1]{Michael Petrov}
\author[1]{Boo Carmans}
\author[1,2]{Josef Soucek}
\author[3]{Akhil Kuriakose}
\author[3]{Ottavia Jedrkiewicz}
\author[1]{Emilie Bourgeois}
\author[1,2]{Milos Nesladek}
\affiliation[1]{Institute for Materials Research, Hasselt University, Wetenschapspark 1, B-3590, Diepenbeek, Belgium}
\affiliation[2]{Faculty of Biomedical Engineering, Czech Technical University in Prague, Sítna sq. 3105, 27201 Kladno, Czech Republic}
\affiliation[3]{Institute for Photonics and Nanotechnology-CNR, University of Insubria, Via Valleggio 11, 22100 Como, Italy}

\begin{document}

\maketitle

\begin{abstract}
    This paper aims to extend the understanding of the mechanism of photo-electrical detection of magnetic resonance (PDMR) in nitrogen-vacancy (NV) centres. This technique is particularly important for development of solid-state quantum computing platforms. In particular, we report on the new insight in the photocurrent (PC) generation and charge cycling in the single NV centre, which is related to PDMR contrast reaching 50\% and above. We develop a technique to locate PC related features. We find that electrons generated at the NV centre are stored in interface trap levels and establish that the interface states serve as an amplifier that can be driven by introducing a second laser into our confocal setup. We show that controlling these interface states allows one to significantly enhance the PDMR contrast. We develop a model that consistently explains observed amplification effects even without the application of a bias voltage.
\end{abstract}
\section*{Introduction}


Optically detected magnetic resonance (ODMR) technique represents a benchmark for reading out nitrogen vacancy spin state in diamond and various other solid state qubits. In 2015, photocurrent-detected magnetic resonance (PDMR) has been introduced,\cite{Bourgeois15} as an alternative spin state readout providing distinct advantages in resolution beyond the limit of confocal optics, important for readout of dipole-dipole coupled electron spins or quantum sensing. Despite these advantages, the photocurrent (PC) is generally less reproducible than photoluminescence (PL). While PL is stable in time, except charge state induced PL blinking, and scales with excitation power in a reproducible manner, the charge carriers, on the other hand, are affected by the defects in the material in which it is generated. In particular they are sensitive to the presence of trap levels and local electric fields. Nevertheless, the PC detection remains an indispensable technique in material research. \cite{Grotz11,Shi15,Ma2023}\\
Nitrogen-vacancy (NV) centre in diamond stands out among other quantum systems thanks to the easy fabrication, accessibility of the spin transition, and long relaxation and coherence times.\cite{Balasubramanian} Their applications range from quantum computing and communication\cite{arute2019quantum, zhong2020quantum,riste2017demonstration,morvan2023phase} to quantum sensing, including magnetometry,\cite{Rondin_2014,Barry20,Wolf15} electric field sensing,\cite{Dolde11,Chen17} thermometry\cite{Kucsko2013,Neumann13} and detection of external electron\cite{Grotz11,Shi15} or nuclear spins.\cite{Mamin13} PL intensity of a negatively charged NV centre (NV$^-$) depends on its spin state. This allows the NV spin state to be read out with the ODMR technique. The quality of the ODMR signal is crucial for all NV centre applications, however, the maximum ODMR contrast is limited by the quantum transition rates of the NV centre and cannot exceed significantly beyond 30\%,\cite{Steiner} or 46\% with the two-power initialization.\cite{Wirtitsch23} This is where PC detection of the NV centre might come in handy.\\
Under laser excitation, NV centres generate free charge carriers by transitioning between negative and neutral charge states. If electrodes are deposited near an NV centre, charge collection is possible, expecting that PC readout depends on NV centre spin states just like PL.\cite{Bourgeois15, Bourgeois17, Gulka17, Bourgeois20, Hrubesch17, Murooka21} We show that unlike ODMR, PDMR contrast is not necessarily limited by the quantum properties of the NV centre and that the effective collection efficiency for current can be much higher than for light. This means a potentially better signal-to-noise ratio (SNR), invaluable for most NV centre quantum applications. An additional benefit of the photoelectric readout is that the device components necessary to perform PDMR can be easier miniaturized. This is relevant for the detection of single NV centres\cite{Siyushev19} as well as for magnetic field sensors based on NV ensembles. However, the PC signal, especially from a single NV centre, is very sensitive to parameters such as diamond surface properties, electrode quality, presence of other defects that act as recombination or trapping centres and material characteristics. Diamonds that contain NV centres usually contain a large number of substitutional nitrogen impurity  N$_\text S$, and possibly also NVH and NVN defect complexes, that generate additional PC under the illumination. This is especially true if NV centres are prepared by high-energy irradiation and annealing. Irradiation, on top of that, can add single vacancy V and double vacancy V$_2$ defects. All these defects can act as recombination centres, which means they make the PC sample dependent.\cite{Bourgeois20} Under certain conditions these defects can also act as photogeneration centres, creating PC background or they can be a source of positive PDMR contrast.\cite{Bourgeois22} However, from the results that we present here, we conclude that there are other factors playing the role in photocurrent processes, in particular charge traps at the interface.\\
The standard way to describe PC from a single defect is through an electric field induced separation of the charge carriers generated at a defect. We shall call this current the inherent NV PC. However, in this study, we find another type of current, the one that is not generated at the NV centre, but at the interface between the electrode and diamond. Yet, this current only exists under the condition that an NV centre is illuminated, and this current also exhibits PDMR contrast. We shall call this current the subsidiary NV PC. The PC that one observes under laser illumination, covering both the NV and the contact interface is then the sum of the inherent and subsidiary NV PC.\\
We have found that PDMR contrast can get as high as 90\% (see SI) in cases when subsidiary NV PC is significant. To characterise this effect we introduce a laser scanning technique, specifically designed for subsidiary NV PC observation. The technique is based on resetting the measurements conditions before each pixel of the scan (see Methods). With this technique, we find that the NV centre influences subsidiary NV PC via trap levels, which can be populated by electrons generated at NV centres. Considering this, we establish a link between the population of the trap levels and high PDMR contrast and come up with a scheme to induce the effect of high PDMR contrast in cases when it is not present from the start. In that case, addition of a second, auxiliary, laser allows us to increase PDMR contrast from 3\% to 20\%. We develop an analytical model to explain the effect of auxiliary laser on PDMR contrast.\\
\begin{figure*}[t!]
    \centering
    \includegraphics[width=0.7\linewidth]{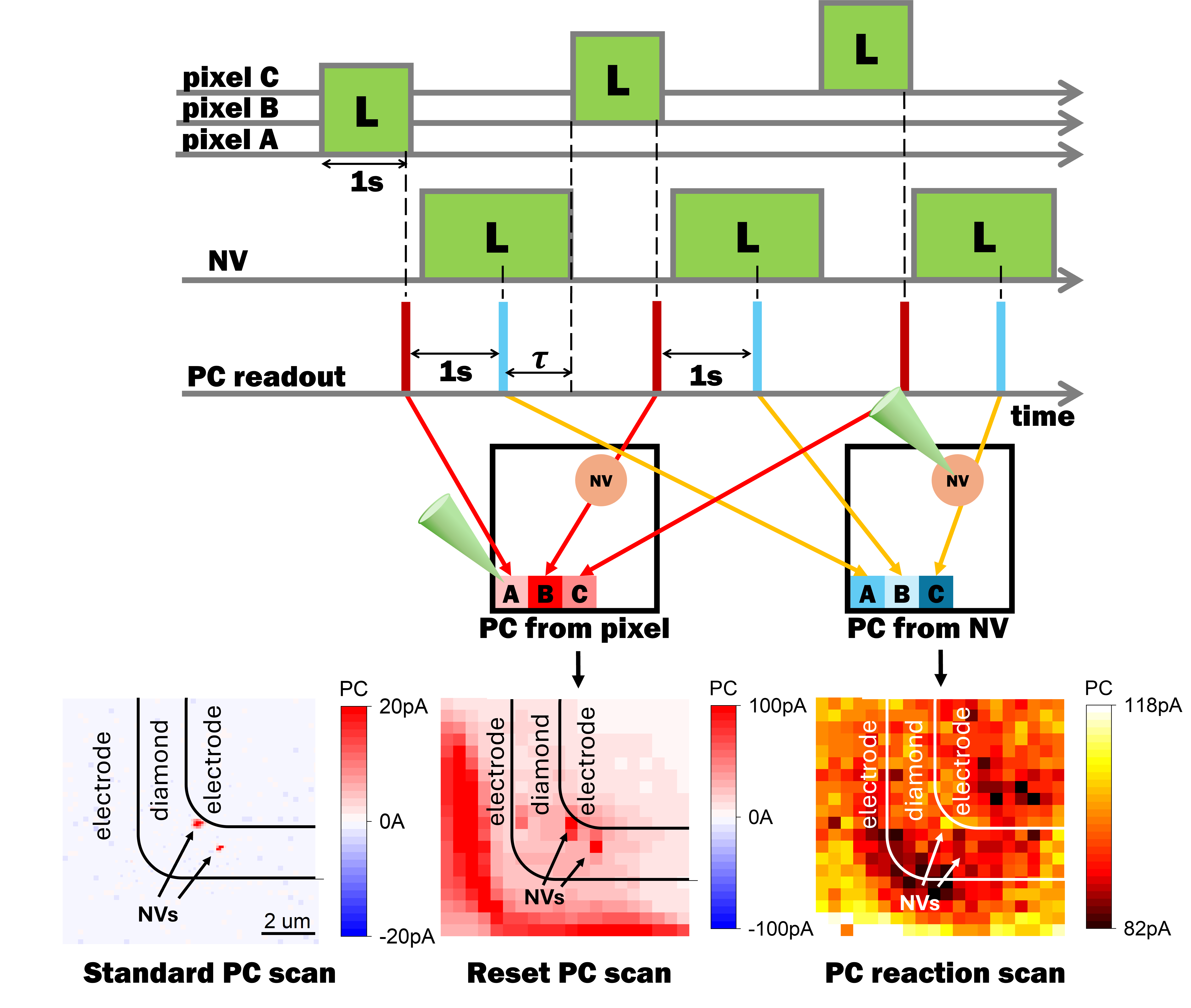}
    \caption{The scheme for photocurrent (PC) mapping used in this work. The position of the laser focus alters between the scanned pixel and the NV centre for which the PC mapping is performed. The PC is recorded one second after the focus is moved to a pixel and one second after the focus moved to the NV centre. The focus is moved to the NV centre immediately after the PC from pixel is recorded. Then the focus is moved to the next pixel in some time ($\tau$) after the PC from the  NV spot is recorded. Laser light is blocked by an acousto-optic modulator (AOM) during focus movement. $\tau$ is not a fixed time, rather it is a time until a certain PC threshold is reached, at which point the measurement conditions for the next pixel are considered to be reset, as discussed in the text. The PC threshold is chosen individually for each NV. Examples of the resulting maps as well as the standard PC map for comparison are demonstrated. We call the map of the PC from pixels - the ’Reset PC scan’, and the map of the PC from the NV spot - the ’PC reaction scan’, discussed in the text.}
    \label{m2}
\end{figure*}
\section*{Methods}
Measurements are performed on a home-built confocal PL microscope equipped with a 561 nm Nd-YAG laser. The laser light is focused onto the diamond surface with a 0.9 NA air objective. Piezo-stage Physik Instrumente P-545 is used for mapping. MW is applied through an antenna which is fabricated together with the electrodes. The distance between the antenna and the working area is 10-20 μm depending on the studied NV centre. The current through the electrodes is measured with Keithley 487 picoammeter (10 fA sensitivity).\\
We have developed a scheme to probe how the PC from the NV spot responds to illumination of different spots on the sample. This scheme is depicted in figure \ref{m2}. This is a regular PC map pixel by pixel with the caveat that between each pixel the focus point moves back to the NV centre. This step resets the measurement conditions for each pixel. This is crucial since the difference between a standard PC scan and a Reset PC scan can be dramatic (fig. \ref{m2}). With this technique we can record the PC from the pixel as well as the PC from the NV spot. Thus, by repeating the scheme for each pixel we are able to obtain two maps.\\
To avoid confusion, we call the map of the PC from pixels after a reset - the 'Reset PC scan', and the map of the PC from the NV spot - the 'PC reaction scan'. If illumination of a pixel has an effect on the PC from the NV spot, this will be visible on the PC reaction scan. These two different types of maps are explained in figure \ref{m2}).\\
When performing such scans, the laser light is blocked by an acousto-optic modulator (AOM) during the movement of the focus point. No MW pulsing is applied. No voltage on the electrode is applied either unless specified otherwise.\\
The measurements are executed by using a 561 nm laser either alone or in combination with an auxiliary 515 nm diode laser. The beam path is aligned with the main 561 nm laser through a dichroic mirror. The auxiliary laser is aligned while monitoring the PC signal, until a signal drop is observed. The focus points of the main and auxiliary lasers are intentionally mismatched.\\
We have used the following two diamond samples in this study:\\
\textbf{Sample M} (metal electrodes sample): IIa HPHT diamond grown by NDT (Saint Petersburg). Nitrogen concentration is 10ppb, boron concentration is 50 ppb. No implantation is performed on the sample. An aluminium layer on top of a 10 nm titanium layer are created by standard lift-off optical lithography. The thickness of the aluminium layer is uncertain.\\
\begin{figure*}[t!]
    \centering
    \includegraphics[width=0.9\linewidth]{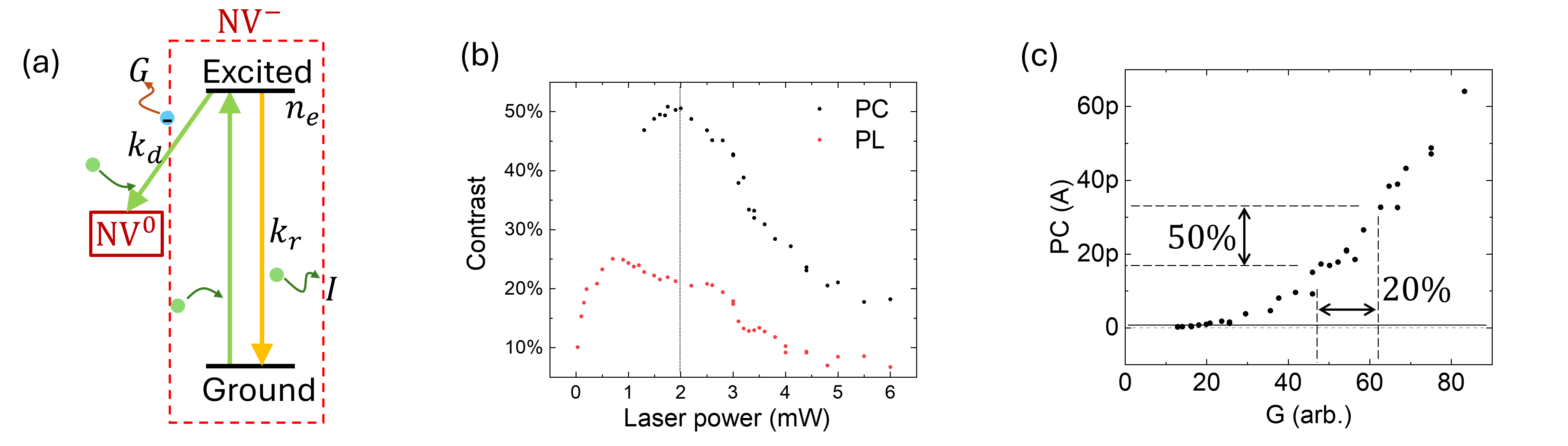}
    \caption{\textbf{(a)} Energy level scheme of NV centre, showing photon absorption, photon generation ($I$), and electron generation ($G$). Only transitions relevant for the discussion are depicted. The green circles are photons, the blue circle is an electron. \textbf{(b)} PC and PL electron spin contrast for different laser powers measured on Sample M. PC contrast repeats the general trend of the PL contrast, while staying significantly higher. \textbf{(c)} PC as a function of the charge generation rate. The dependence is non-linear, which leads to the difference in PL and PC contrast. $20\%$ ODMR contrast translates to $20\%$ contrast in $G$, which, in turn, translates to $50\%$ PDMR contrast. In (b) such ratio between ODMR and PDMR contrast is observed at 2 mW laser power.}
    \label{nl}
\end{figure*}
\textbf{Sample G} (graphite electrodes sample): IIa HPHT diamond grown by NDT (Saint Petersburg), implanted by DIATOPE (Ulm). Nitrogen concentration 10ppb, boron concentration is 50 ppb. Implantation parameters are 5 keV acceleration energy and $5*10^8 \frac{^{15}\text N^+}{\text{cm}^2}$ ion dose, with subsequent annealing at 1000$^\circ \text C$ for 3 hours in vacuum. Graphite electrodes are created on the sample by IFN-CNR (Como). The graphite electrodes are fabricated using a Bessel laser beam, whose shape enables the formation of graphitic structures extending through the whole sample thickness. The graphite electrode is a graphite column orthogonal to the sample surface. The diameter of the column is 5 μm. The column extends through the entire thickness of the sample. In order to connect the graphite columns to an ammeter, a 100 nm aluminium layer on top of a 20 nm titanium layer are created by standard lift-off optical lithography.\\

\section{Charge carrier generation rate G}
\label{ccgr}
We define the charge carrier generation rate ($G$), which is the number of charge carriers produced by NV per unit of time. $G$ depends only on the NV transitions (fig. \ref{nl}a):
\begin{equation}
    G=2n_ek_d
    \label{nekd}
\end{equation}
where $n_e$ is the population of the excited state and $k_d$ is the transition rate NV$^-$$\rightarrow$NV$^0$ (which we denote as deionization, considering that NV$^-$ becomes electrically neutral as a result). The factor of two comes after considering both electrons and holes generated at NV centre. In (\ref{nekd}) and in figure \ref{nl}a, we assume that the dominant transition path for the deionization is the electron transition from the NV$^-$ excited state to the conduction band.\cite{Gali19} Since NV deionization requires a photon, $k_d$ is proportional to laser power $P_\text{NV}$. Considering also that $I\propto n_ek_r$, where $I$ is the PL intensity and $k_r$ is the radiative transition rate, which is a constant, we conclude that 
\begin{equation}
   G\propto IP_\text{NV}
   \label{G}
\end{equation}

\begin{figure*}[t!]
    \centering
    \includegraphics[width=0.8\linewidth]{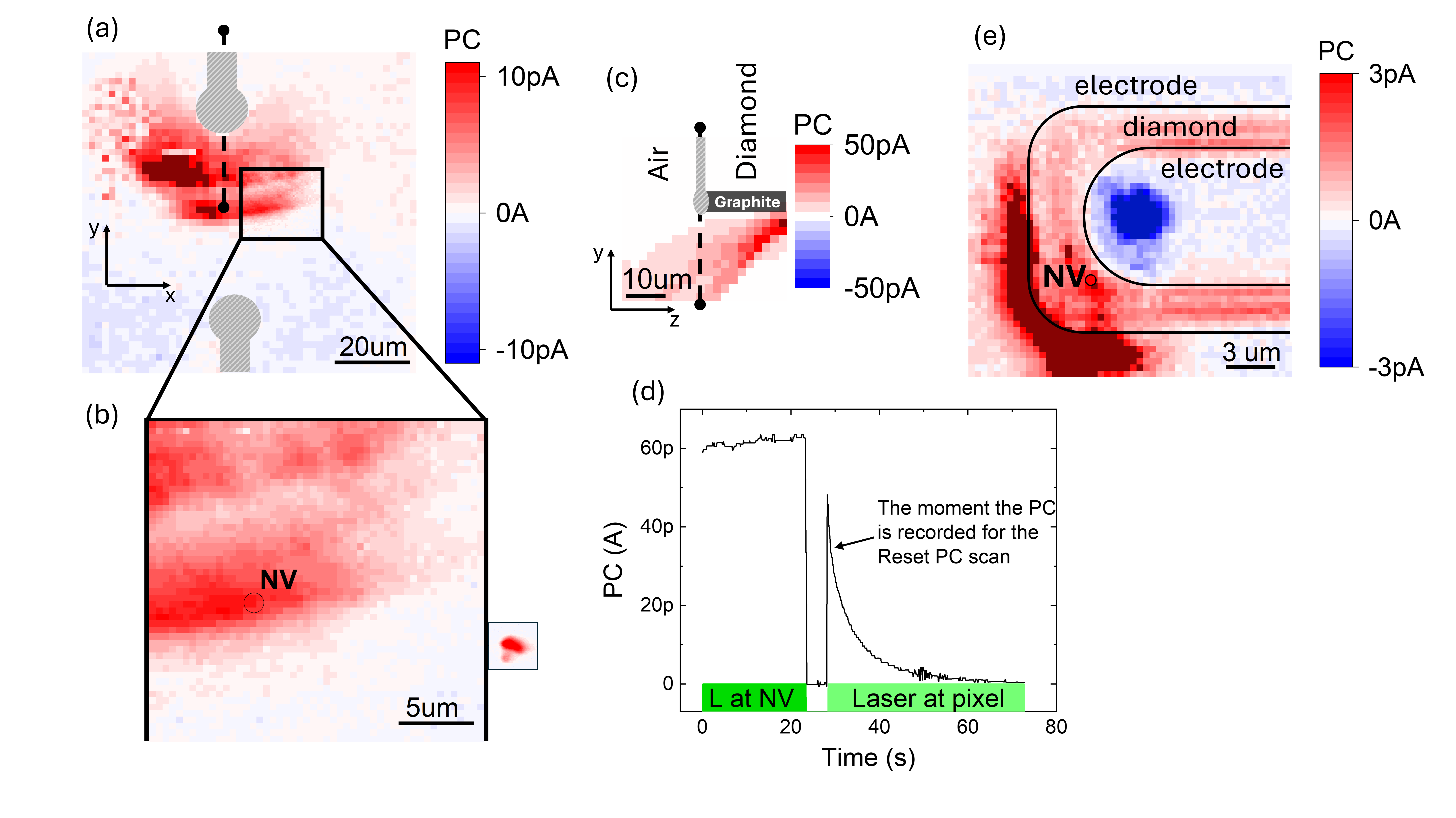}
    \caption{\textbf{(a)} The Reset PC scan of the surface (xy-scan) around the electrodes of the Sample G. \textbf{(b)} The Reset PC scan of the surface (xy-scan) around the NV centre. No additional PC is detected with NV in focus compared to the PC in the vicinity. As a reference, the inset depicts the result of a standard PC scan around the NV centre. Multiple single NV centres are in the vicinity, which makes the map look irregular. \textbf{(c)} The Reset PC depth scan (yz-scan, where z is the coordinate perpendicular to the diamond surface) around a graphite electrode. The dashed lines here and in (a) mark the intersection of these two perpendicular scans. The PC is originating form a single point 50 μm below the surface (after correction for refraction). Note that due to the cone shape of the beam, single points appear as cone shapes on depth scans (see SI). \textbf{(d)} The PC as a function of time before and after illumination of a pixel in (a), (b) or (c). The PC decays to 0 pA after 30 seconds. The NV would need to be illuminated again to restore the PC from the pixel.}
    \label{bs}
\end{figure*}
This allows us to treat $G$ as an experimentally observable parameter. From (\ref{G}), it follows that $G$ must have the same microwave (MW) contrast as PL signal. Yet, the PDMR and the ODMR contrast can differ dramatically (Fig. \ref{nl}b). Further on, we consider the relation between $G$ and PC. We find that PC ($J$) in the case of NV centres presenting a high PDMR contrast is not a linear function of $G$ (fig. \ref{nl}c). A decrease in $G$ is due to the MW resonance and it translates to a decrease in $J$ according to the dependence $J(G)$ (see SI). The origin of this non-linearity will be explained in sections \ref{toot} and \ref{tiec}.\\

\section{The origin of the PC}
\label{toot}
\begin{figure*}[t!]
    \centering
    \includegraphics[width=0.6\linewidth]{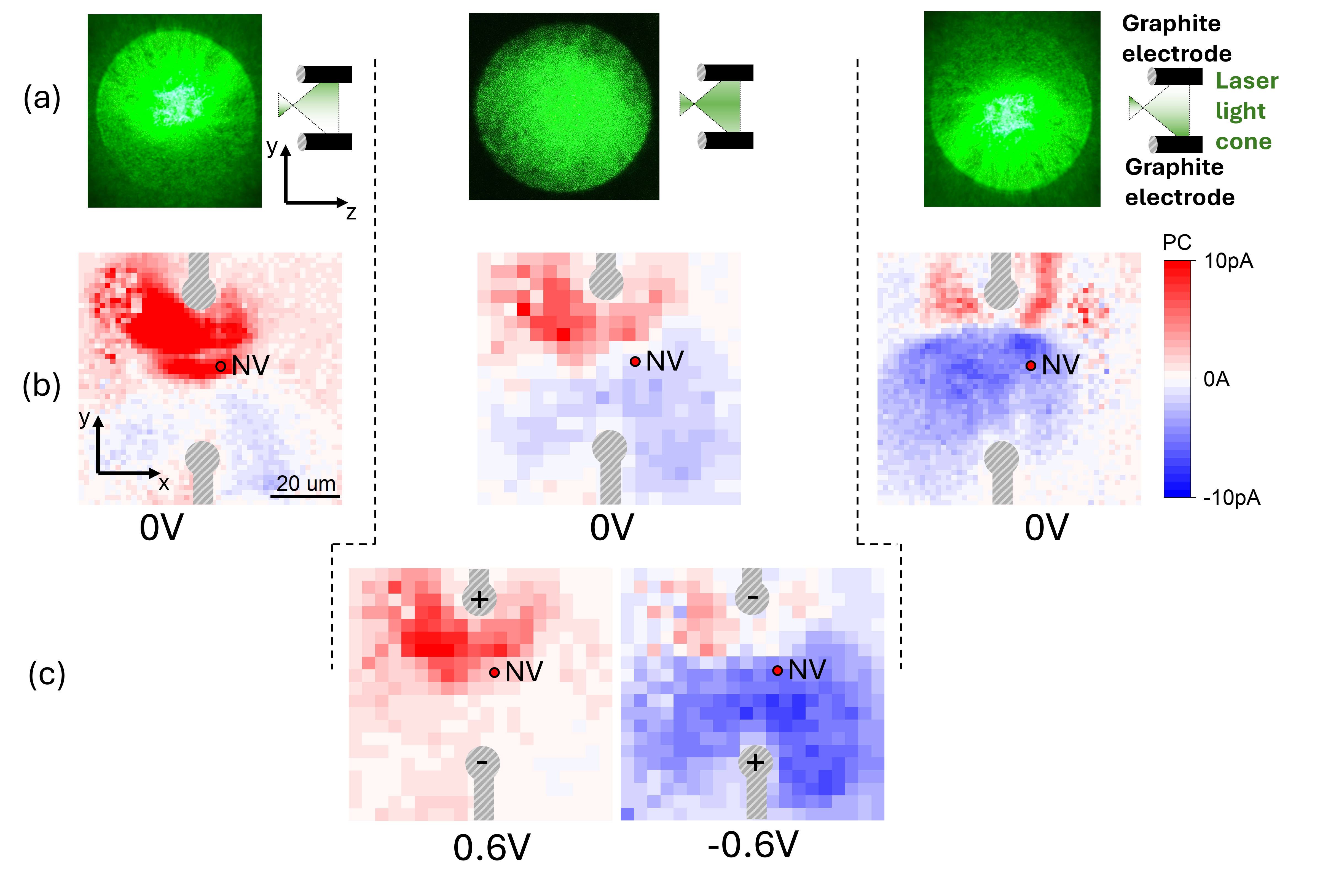}
    \caption{\textbf{(a)} Beam shapes after optical objective and the corresponding graphite electrode illumination pattern. \textbf{(b)} Reset PC scans of the sample surface (see Methods) with respective beam shapes from (a). The scans are made under 0V bias. The electrodes on the surface are marked grey. The scans with the antisymmetric beam profile show that the electrode, which is illuminated stronger exhibits a stronger PC profile around it. \textbf{(c)} Reset PC scans with symmetric beam profile under bias voltage. The general pattern shape is preserved, while a PC offset is added in the direction of the voltage.}
    \label{shp}
\end{figure*}
We perform the Reset PC scans on the Sample G.  We see that a large area of the sample surface exhibits a PC response (fig. \ref{bs}a). However, this PC is only visible because the NV (marked in fig. \ref{bs}b) is illuminated prior to each pixel (see Methods). This means that NV centre illumination creates the conditions for the PC to flow, which are only satisfied for a short time (in the range of tens of seconds, fig. \ref{bs}d). 
The Reset PC depth scan (fig. \ref{bs}c) shows that the PC that we observe when scanning the surface is originating from a single point on a graphite electrode, i.e. the PC is highest when the laser is focused on that spot. We call this point a Source. This Source is about 50 μm deep at the electrode-diamond interface, after correcting for light refraction (see SI). There is also a second Source found at the other graphite column (not shown). It is possible to see the PC from the Source without NV illumination, i.e. only the Source must be illuminated, however the PC is two orders of magnitude smaller if NV is not illuminated (see SI). This clearly confirms the NV involvement in the generation of current from the Source. We anticipate that there is a mechanism in which illumination of an NV centre amplifies the PC from the Source for a limited time. This explains why, when focusing outside of NV centre during a standard PC scan, we see no PC, unless we have illuminated an NV centre directly before that (which is what we do for every pixel during Reset PC scans, fig. \ref{bs}d).\\ 
In order to distinguish between the PC from a Source (Source PC) and the PC from NV (NV PC), we compare the PC at the point of the NV centre and the PC in its immediate vicinity for Sample G. We expect that exactly at the point of the NV centre the total PC would be significantly higher, which is something we detect regularly with Sample M (fig. \ref{m2}). However, the scan reveals that this is not the case for Sample G. For Sample G, there is no detectable increase in PC when the focus point is exactly on the NV centre (fig. \ref{bs}b). This means that the NV PC in this case is negligible compared to the Source PC. A standard PC scan, for comparison, still shows that the PC is only present when the focus point is exactly on the NV centre (fig. \ref{bs}b-inset). \\
Since the PC that we observe while performing the Reset PC scan of the surface is actually due to charge injection at the two Sources approximately 50 μm below the surface, we expect that the shape of the pattern visible in figure \ref{bs}a is related to the laser beam profile. Indeed, when we change the beam profile, as depicted in figure \ref{shp}, we see that it has a dramatic effect on the PC at the sample surface (fig. \ref{shp}a,b). This indicates that the beam trajectory through the crystal towards the Source, as depicted in SI, is very important. Larger PC is observed at the electrode, which is illuminated stronger. The shape of the Source PC at the surface is then dependent on how efficiently the two Sources are illuminated at each point of the scan. The two Sources produce current in opposite directions and the direction of the total current is determined by the relative illumination efficiency of the two Sources.\\
We expect that finding the Sources will be more difficult for Sample M as most of the electrode-diamond interface of Sample M is masked by the electrodes, which makes it difficult to access with laser light. However, it seems that despite that, we can still observe the Sources of Sample M. The reset PC scan of the surface of Sample M shows that the PC is highest when laser light is focused directly on top of the electrodes (fig. \ref{bs}e). The PL spectra measured at the electrodes (see SI) show diamond Raman peaks, which suggests their transparency, therefore the aluminium must be not thicker than a few nm. This means that laser light can reach the diamond-metal interface even through the electrodes. We conclude that the areas of high PC (fig. \ref{bs}e) are the Sources of Sample M.\\
We perform the scans with and without bias voltage (fig. \ref{shp}b,c). Voltage application results in a PC offset, which is almost universal across the surface. The general pattern of the PC response remains unchanged as voltage is applied. From the direction of the PC, we learn that a Source must be either a point where holes are injected into diamond, or a point where electrons are drained from diamond. The latter implies presence of intrinsic free charge carriers in diamond, and therefore we discard this option. This leaves the hole injection. We assume the following for the hole injection mechanism: The holes are created at a Source by a non-luminescent diamond defect which exists exclusively on the metal interface, and the holes are separated with electrons by the local electric field of the depletion region. It still remains unclear why Sources are localized to spots on a diamond-electrode interface and are not spread over the entire surface. We would like to mention, however, that the locations of the Sources on Sample G seems to be related to the illumination pattern (see SI)\\
\subsection*{Summary}
In Sample G and Sample M, the PC is not directly a result of charge carrier generation at single NV centres. The PC is coming from certain locations on the diamond-metal interface, which we call Sources. The charge carriers produced by the NV centre act as an amplifier for Source PC. In Sample M, the Sources are located under the electrodes and could be accessed with laser through partially transparent electrodes. Though the exact mechanism is unknown, it is clear that a Source is a point where holes enter diamond.

\begin{figure*}[t!]
    \centering
    \includegraphics[width=\linewidth]{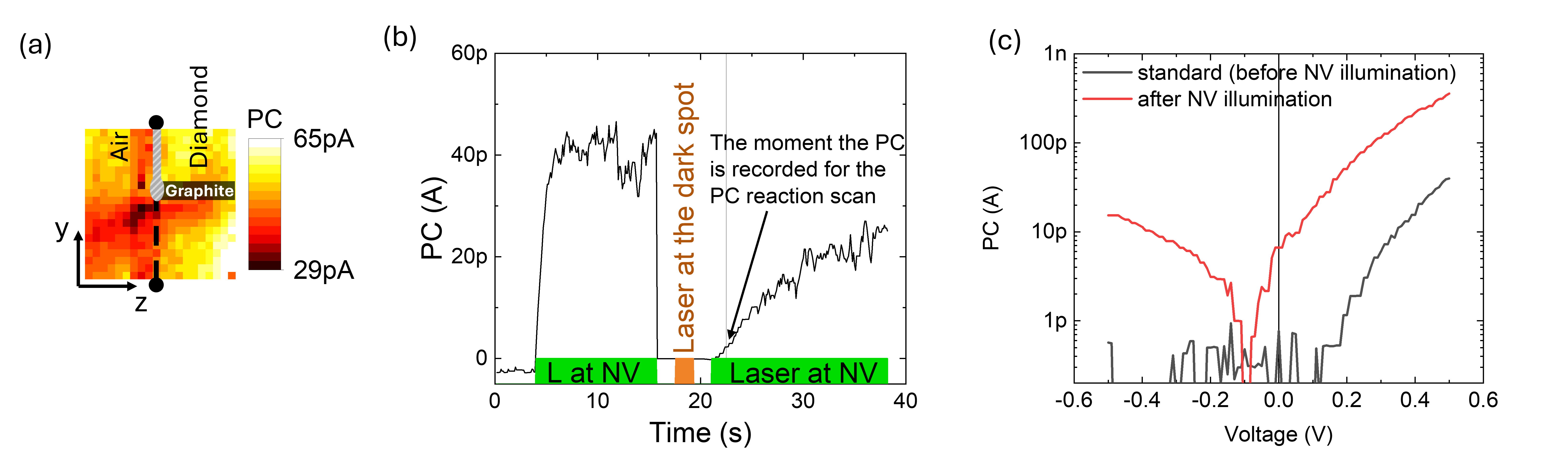}
    \caption{\textbf{(a)} The PC reaction depth scan (yz-scan) around a graphite electrode. The PC drops significantly after illuminating the intersection of diamond, electrode and air, which results in the dark spot in that area of the scan. \textbf{(b)} The PC as a function of time before and after illumination of the dark spot in (a). The NV PC drops significantly after the illumination of the dark spot but slowly rises back once the NV is back in focus. \textbf{(c)} Current-voltage characteristic of a Source before and after illumination of NV centre. Laser power is 3mW. Bridge population is supposedly zero before NV illumination and non-zero after illumination.}
    \label{briv}
\end{figure*}
\begin{figure*}[t!]
    \centering
    \includegraphics[width=0.7\linewidth]{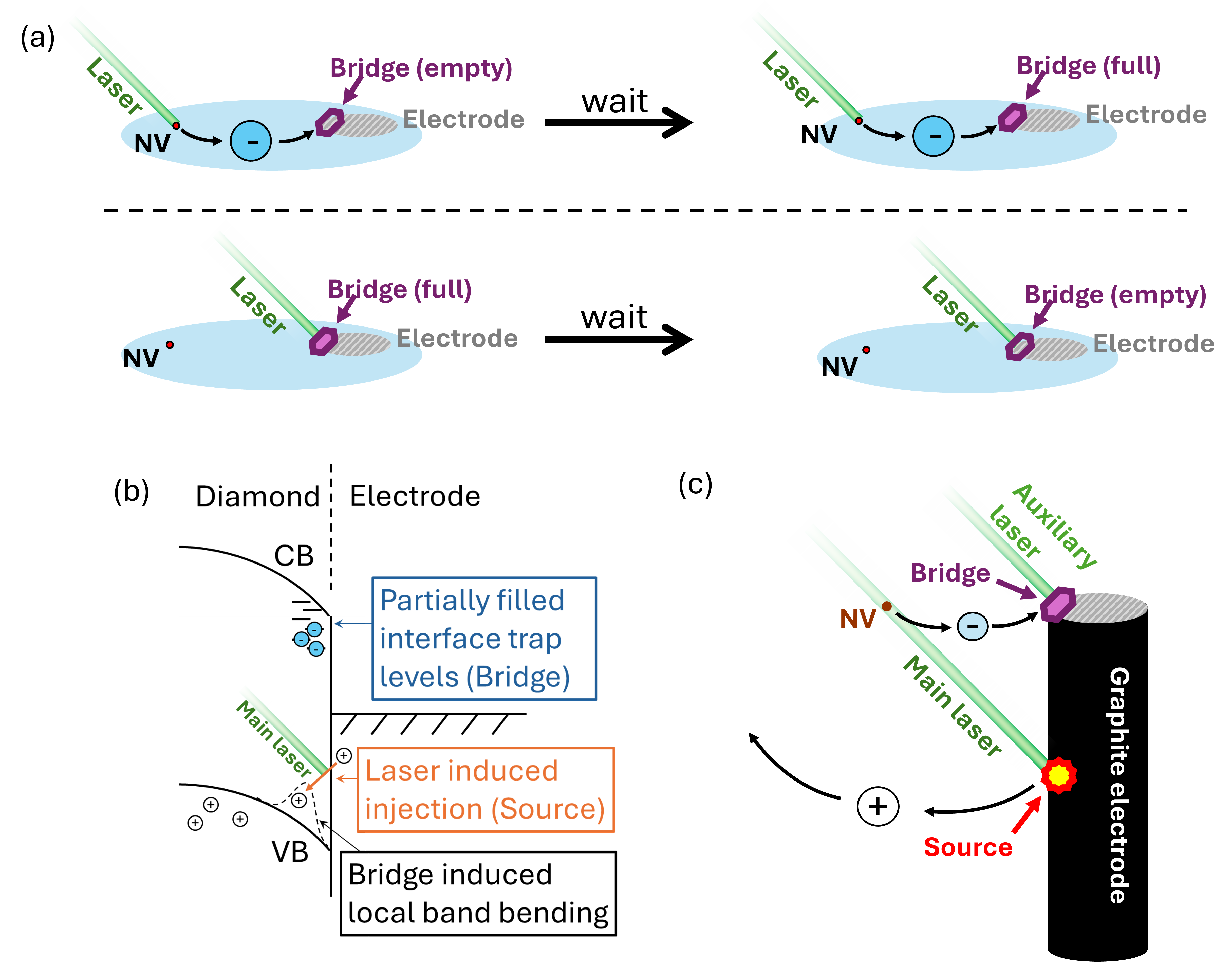}
    \caption{\textbf{(a)} Suggested Bridge cycle. NV generates electrons under laser illumination. These electrons travel by diffusion to the Bridge and populate it. When illuminated, the Bridge loses trapped electrons to the electrode and gets depleted. \textbf{(b)} A graphical description of the Bridge and the Source on a band diagram of the diamond-electrode interface. The case for a reverse Schottky barrier is depicted based on high boron concentration in diamond (see Methods). The potential barrier for holes is altered when the Bridge is populated, which increases the efficiency of hole injection at the Source. \textbf{(c)} The suggested process for PC generation in the presence of the Aux laser. NV centre under illumination generates electrons, which populate trap levels at the Bridge, while the Aux laser depopulates the trap levels at the Bridge. The Source under illumination injects holes into diamond. These holes are the only considerable contribution to the PC. The number of injections per unit time depends on the Bridge population.}
    \label{in}
\end{figure*}

\section{The intermediate effect connecting an NV and a Source}
\label{tiec}
Figure \ref{briv}a shows that illumination of a certain area at the surface near the electrode results in a decrease of Source PC. The area appears to be at the intersection of electrode, diamond and air. We call this area a Bridge. It can be also found that the PC can be restored back to its original value by illuminating the NV centre for some time (fig. \ref{briv}b). Our hypothesis is that the Bridges consist of trap levels, which are populated with charge carriers generated at NV centres. The traps, then, can also be depleted with light illumination (fig. \ref{in}a). We convey this hypothesis in the following formula:
\begin{equation}
    \dot B= G^a\left(B_\text{max} - B\right) - P_\text B^bB - n_\text{rec}B
    \label{b0}
\end{equation}
where $B$ is the population of the trap levels (which we will call the Bridge population), $G$ is the charge carrier generation rate from the NV centre as introduced in section \ref{ccgr}, $P_\text{B}$ is the number of photons per unit time (i.e. laser power) of Bridge illumination, $B_\text{max}$ is the Bridge population when all trap levels are full, $n_\text{rec}$ is the rate of creation (i.e. generation at defects and, possibly, injection) of the type of free charge carriers which depopulates the trap levels via recombination. We are omitting the constant coefficients, such as capture cross sections, and are using unitless parameters for simplicity. We expect that $a=1$ since a single charge carrier is captured by a trap independently of other charge carriers (superposition principle). Parameter $b$ could be either $b=1$ or $b=2$, depending on the number of photons required to depopulate the trap. The lasers we use have photon energies of 2.21 eV (main laser) and 2.42 eV (auxiliary laser). In both cases it is likely that a single photon is enough to depopulate the trap, so we expect that $b=1$.
\\ 
We observe that Source PC depends on the Bridge population (fig. \ref{briv}a). While populated, the traps produce electric field which spreads to a Source (fig. \ref{in}b). We anticipate that the local bend bending introduced is going to lower the energy barrier for holes. This, in turn, could lead to the amplification of Source PC.  Additionally, in this process, the NV is needed to produce charge carriers which change the occupation at the Bridge. We have observed that Bridge population increases Source PC by approximately the same factor for any voltage (fig. \ref{briv}c). Therefore we have chosen to approximate Source PC with the first order polynomial of $B$:
\begin{equation}
    J=(B^c+1)P_\text{S}^d J_\text{S}(U)
    \label{j1}
\end{equation}
where $J$ is Source PC (which we consider to be the same as the total PC), $P_\text{S}$ is the number of photons per unit time (i.e. laser power) of Source illumination, $J_\text{S}(U)$ is the current-voltage characteristic of a Source when $B=0$ (fig. \ref{briv}c). Parameter $c$ depends on the mechanism of the interaction between a Bridge and a Source and is difficult to estimate since we cannot measure $B$ directly. With respect to parameter $d$, we expect that either $d=1$ or $d=2$, depending on whether the charge carrier creating process at a Source is a one- or two-photon process. For a detailed derivation of (\ref{b0}) and (\ref{j1}) see SI.\\
From (\ref{b0}) and (\ref{j1}) it follows that if $G$ and $P_\text B$ are kept constant, $J$ would reach the steady state $J_\text{st}$:
\begin{equation}
    J_\text{st}= \left[\left(\frac{G^aB_\text{max}}{G^a+P_\text B^b+ n_\text{rec}}\right)^c+1\right]P_\text{S}^d J_\text{S}(U)
    \label{j2}
\end{equation}
This explains the non-linearity in the function $J(G)$ discussed in section \ref{ccgr}. This non-linearity, in turn, is responsible for the significant difference between ODMR and PDMR contrast, i.e. the effect of high PDMR contrast.\\
According to our hypothesis, an increase in Bridge population leads to an increase in Source PC, however, we do not observe any PC increase while illuminating a Source alone (see SI). In fact, the only time we observe Source PC amplification is when we illuminate an NV centre. From this we conclude that the Bridge (meaning trap levels of the Bridge) can only be populated as a result of NV illumination, and that the type of charge produced at the Source cannot populate the Bridge. We observe that Bridges tend to move towards positive field bias (see SI). This makes us believe that Bridges consist of electron traps. A Source would then be a point where holes are injected into a diamond, which agrees with a conclusion from the previous section.\\
Considering the latter conclusion, the rate $n_\text{rec}$ of creation of charge carriers which can recombine with the traps, depopulating them as a result, can be expected to be dependent primarily on the number of charge carriers injected at the Source, i.e. Source PC:
\begin{equation}
   n_\text{rec}\propto J_{st}
   \label{pse}
\end{equation}

\subsection*{Summary}
The surface area producing  the PC drop, i.e. the Bridge, consists of electron trap levels, which can be populated with charge carriers from NV centre, and depleted with light (fig. \ref{in}a). Bridge population greatly affects Source PC, presumably via generation of the local electric field and its alteration At a Source holes, not electrons, are injected (fig \ref{in}b,c).

\begin{figure*}
    \centering
    \includegraphics[width=\linewidth]{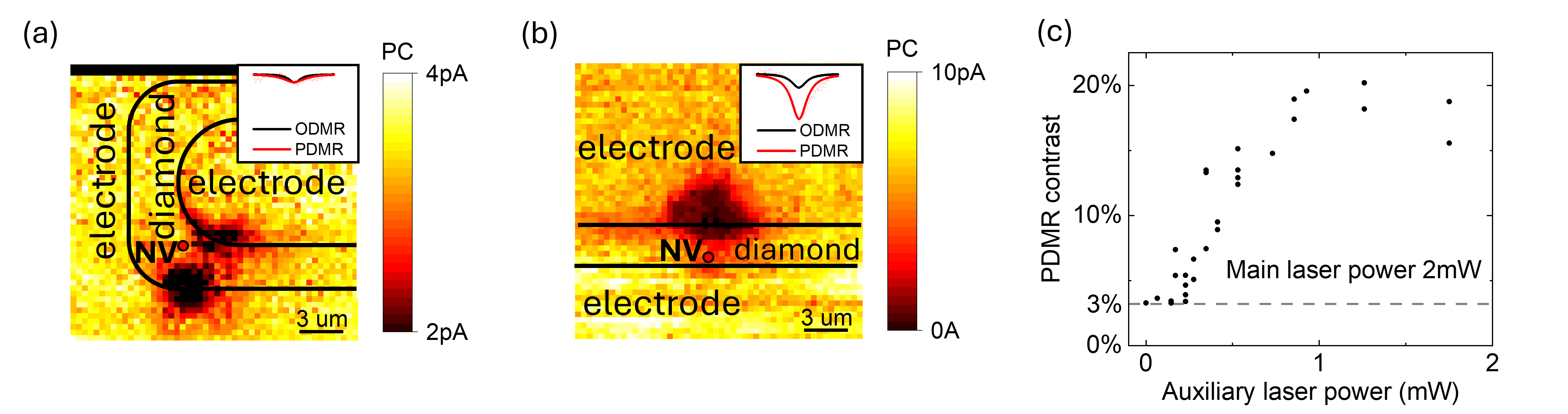}
    \caption{\textbf{(a)} The Reset PC scan of the surface of the Sample M. The PC is highest when illuminating the electrodes. \textbf{(b)} The PC reaction scan of the same area as (a). There are two spots on the scan. The PDMR contrast at the NV used for the scan is 9\% (inset). \textbf{(c)} Same as (b), but around a different NV centre. There is only one spot and the PDMR contrast is 50\%. More examples of the correlation between the number of spots and the PDMR contrast are in SI. \textbf{(d)} PDMR contrast in presence of auxiliary illumination. With the auxiliary laser, it is possible to increase contrast 7 times.}
    \label{hc}
\end{figure*}
\section{PDMR contrast enhancement}
According to (\ref{j2}),
\begin{equation}
    \lim_{G^a\gg P^b_\text B + n_\text{rec}}J_\text{st}= \left[B_\text{max}^c+1\right]P_\text{S}^d J_\text{S}(U)
    \label{j2}
\end{equation}
here $J_\text{st}$ is not a function of $G$, hence, no PDMR contrast is possible. From this formula, it follows that, according to our model, Bridge illumination ($P_\text B$) must greatly affect PDMR contrast. We confirm this with measurements on Sample M.
Bridges of Sample M are more localized than those of Sample G. Only two spots of 2-3 μm diameter, one at each electrode, appear to have the effect of the PC drop (fig. \ref{hc}a). In some instances, only one spot is visible (fig. \ref{hc}b). We believe the reason for that is the proximity of the NV centre to one of the electrodes. Whenever there is only one spot visible, the NV centre is close to one of the electrodes (see SI). Most likely, the second spot cannot appear after illumination of the Bridge area because the Bridge is illuminated already every time when the NV centre is in focus, which happens after every pixel. The focus spot size in our measurements is about 0.3 μm, which is often larger than the distance between an NV centre and an electrode, therefore simultaneous illumination of an NV and a Bridge seems plausible.\\
Since an NV centre serves as an amplifier for Source PC, the effect of NV spin resonance can also be visible in Source PC. However, the PDMR contrast measured on the Source PC rarely exceeds 3\% unless a Bridge is illuminated simultaneously with NV centre. The PC reactions scans of Sample M reveal a correlation between the PDMR contrast and the number of dark spots visible on the scan (fig. \ref{hc}a,b, see SI for more examples). NV centres which are located close to a Bridge exhibit high PDMR contrast of 20\% and above. When a PC reaction scan is made using such an NV centre, only one dark spot appears on the scan. This correlation can be explained considering the Bridge behaviour and its effect on Source PC. Source PC depends on the Bridge population, Bridge population goes up when NV centre is producing electrons and goes down when the Bridge is illuminated. This means that if the Bridge is not illuminated, the Bridge population will saturate and will not depend on NV centre anymore.\\
Considering that Bridge illumination is important to achieve the high PDMR contrast, we have added an auxiliary laser to our optical setup. The purpose of the auxiliary laser is to illuminate the Bridge, while the main laser illuminates the NV centre and the Source simultaneously (fig. \ref{in}c). Indeed, measurements of the PDMR contrast as a function of the auxiliary laser power (fig. \ref{hc}c) have shown that by adding an auxiliary laser it is possible to increase PDMR contrast from 3\% to 20\%.\\
\subsection*{Summary}
When the Bridge is illuminated simultaneously with NV centre, the PDMR contrast is high. PDMR contrast can be enhanced with additional laser.\\

\begin{figure*}[t!]
    \centering
    \includegraphics[width=0.7\linewidth]{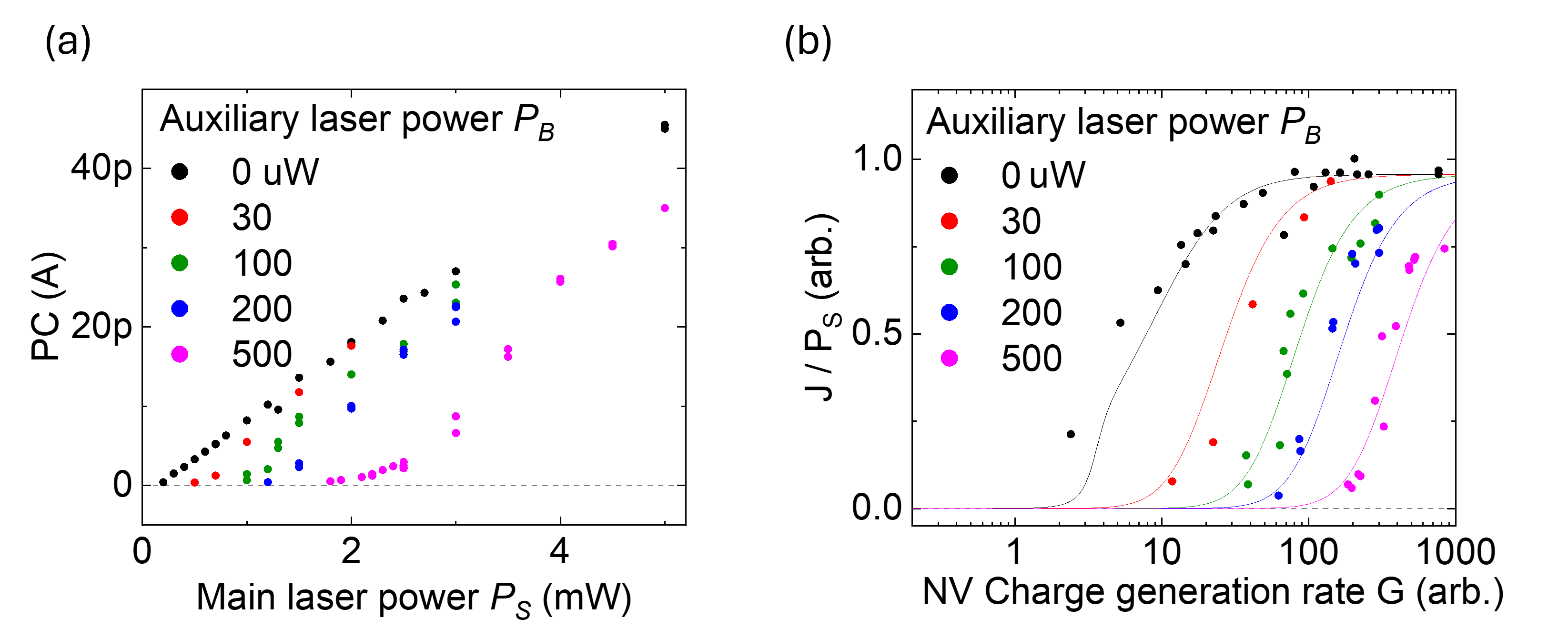}
    \caption{\textbf{(a)} PC as a function of the main laser power for different auxiliary laser powers. Without auxiliary laser, the dependence is linear. Addition of auxiliary laser introduces the non-linear element into the dependence. \textbf{(b)} Ratio between PC and laser power as a function of the main laser power. The curves are fit with the model (see main text).}
    \label{mod}
\end{figure*}

\section{Parameters estimation and model test}
We will use the experimental data presented in figure \ref{mod}a to find the values of the free parameters $a$, $b$, $c$, $d$ in (\ref{j2}) and ultimately test our model. The experiment was performed at $U=0$ and we observe that $J_\text{st}(U=0, G=0)\approx0$ (fig. \ref{briv}c). Considering (\ref{pse}), this allows us to simplify (\ref{j2}):
\begin{equation}
    J_\text{st}= \left(\frac{G^aB_\text{max}}{G^a+P_\text B^b+ J_\text{st}}\right)^cP_\text{S}^d J_\text{S}(0)
    \label{j4}
\end{equation}

The power of the auxiliary laser corresponds to $P_\text B$, while the power of the main laser corresponds to $P_\text S$. The main laser is also responsible for NV illumination ($G$). Since we do not observe PC saturation at high laser powers (fig. \ref{mod}a), we assume that for $G\rightarrow\infty$ relation $G^a>J_\text{st}$ is satisfied. Then, we can find parameter $d$ by calculating $J_\text{st}$ at the limit of high main laser power, which is also the limit of high $G$:
\begin{equation}
    \lim_{G\rightarrow\infty} J_\text{st}= B_\text{max}^cP_\text{S}^d J_\text{S}(0)
    \label{j5}
\end{equation}
According to the experimental data (fig. \ref{mod}a) $J_\text{st}$ is the linear function of laser power at the limit of high laser powers. Therefore we conclude that $d=1$ and the charge carrier creating process at the Source is a one-photon process. With this, we fit the datasets presented in figure \ref{mod}a. We get the parameter values $a=2$, $b=2$ and $c=2$, and the following function as the result:
\begin{equation}
    J_\text{st}= \left(\frac{G^2B_\text{max}}{G^2+P_\text B^2+ J_\text{st}}\right)^2P_\text{S} J_\text{S}(0)
    \label{j7}
\end{equation}
The result of fitting the experimental data with this function is presented in figure \ref{mod}b. The function suggests that Bridge depopulation is a two-photon process.
\subsection*{Summary}
The suggested model for the relation between NV centre and Source PC correctly describes PC behaviour at the investigated parameter range. Experiments at different conditions would enable to confirm the validity of the model.


\section*{Conclusion}

We have demonstrated that the charge carriers generated at an NV centre populate the traps at the diamond-electrode interface. We define points on the diamond-electrode surface that we call as the Source and Bridge, at which the charge exchange is tightly related to the carrier generation at the NV centre. We observe a very high PDMR contrast as a result of the interaction between Source, Bridge and NV centre.  The effect of high PDMR contrast in single NV centres is related with the effect of PC drop after illumination of a Bridge. We attribute this PC drop to interface trap levels.\\
We define specific procedure of subsequently illuminating  the NV centre and the electrode in order to quantitatively characterise the subsidiary PC. We show the relevance of subsidiary NV PC and argue that the effect of high PDMR contrast is pronounced only in subsidiary NV PC, and not in inherent NV PC. In one of the samples we were able to find that subsidiary NV PC originates at a point on the diamond-electrode interface, called the Source, which could be an effect of light-assisted charge injection or due to charge generating defects at the interface.\\
Based on these results, we were able to increase the PDMR contrast from 3\% to 20\%, by adding an auxiliary laser to our optical setup. Introducing control of the auxiliary laser focus position would be the next step in improving PDMR contrast even further.\\
We develop a model to explain the contrast increase and the PC drop. The model is based on the assumption that trap levels at the Bridge control the injection current. We approximated phenomenologically  the injection current based on the observed PC behaviour. The model suggests that laser induced trap level depletion is a two-photon process. This preliminary model was used to explain the relation between the trap levels population and the high PDMR contrast. Comparison between the model and experimental data suggests that the laser-induced trap level depletion is a two-photon process. Data from more samples with different electrode configurations would enable to build a more complete picture of the processes, which are responsible for Source PC and high PDMR contrast.

\section*{Acknowledgement}
We acknowledge the support from FWO (Funds for Scientific Research) Flanders, Projects No. G0D1721N and No. G0A0520N and project EOS CHEQs, EU No. 101046911 (QuMicro), Eu -Quantera project QuMaestro and SBO project DeQuNet.\\
We acknowledge the contribution of Michal Gulka, who was the first to observe the effect of high PDMR contrast.

\bibliographystyle{unsrt}
\bibliography{lit}
\end{document}


\maketitle
\section*{Experimental Setup}
\begin{figure}[t!]
    \centering
    \includegraphics[width=0.8\linewidth]{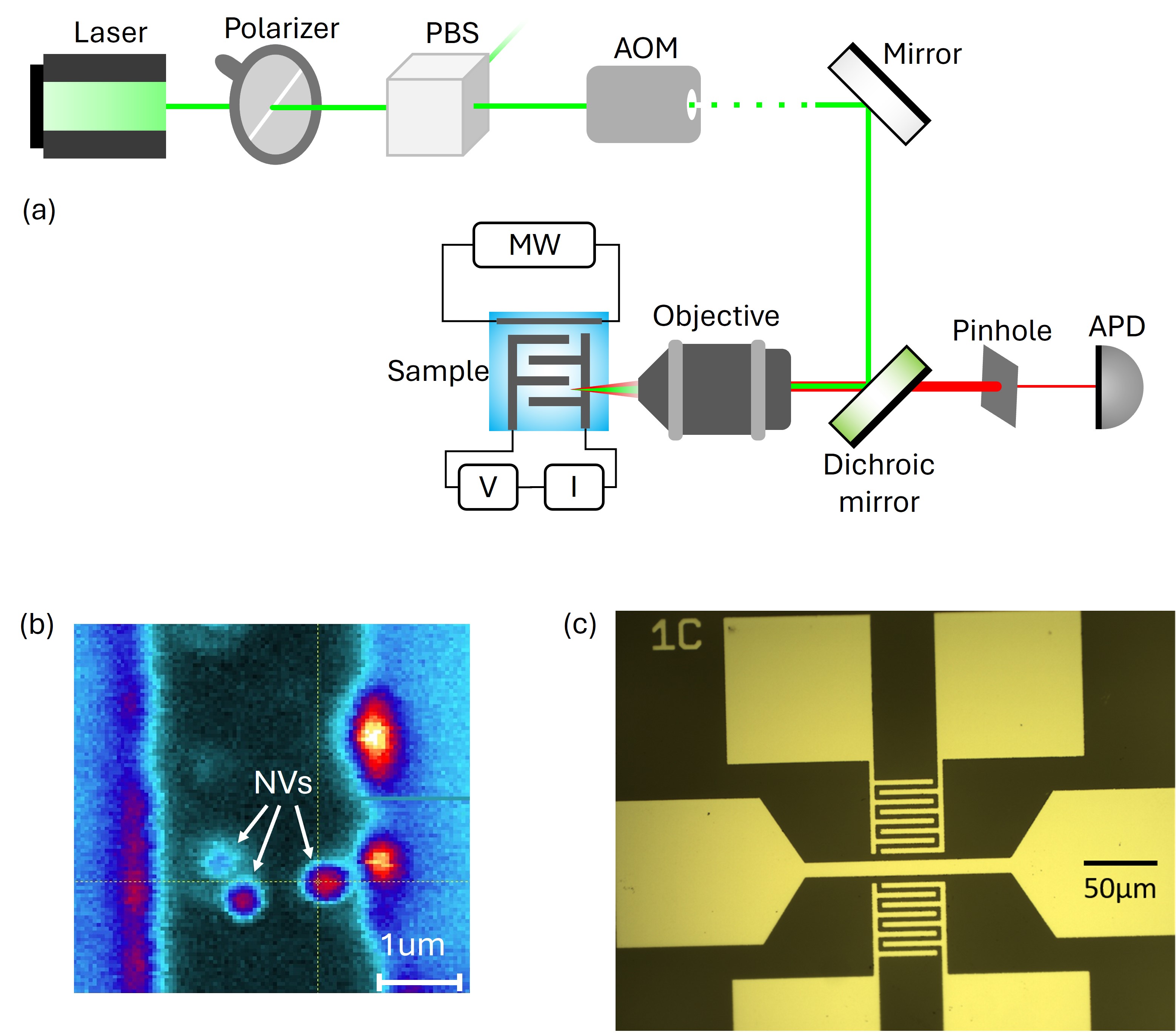}
    \caption{\textbf{(a)} Schematic overview of the measurement setup. Standard confocal setup with the possibility to apply microwaves. Laser goes through a polarizer half-wave plate and a polarizing beam splitter (PBS) used for power control, and an acousto-optic modulator (AOM) to the sample, illumination from the sample is collected through wavelength filters to an avalanche photodiode (APD). A dichroic mirror is used to combine sample irradiation with light collection. \textbf{(b)} A typical photoluminescence map recorded on a single NV sample. NV centres can be clearly seen. \textbf{(c)} Titanium-Aluminum electrodes and microwave strip deposited on diamond surface. Electrodes either above or below the strip are connected to the electric circuit. One end of the strip is connected to a microwave amplifier, the other to a 50 Ohm load. Optical microscope image.}
    \label{setup}
\end{figure}

Figure \ref{setup}a depicts a standard quantum confocal setup used for measurements. The experiments start with the initialization of the spin system by optical pumping to the state $m_s=0$ using a yellow 561 nm continuous wave, solid-state Gem 561 laser manufactured by Laser Quantum. The laser light passes a WPHSM05-561 half-wave plate from Thorlabs and a PBS201 -20 mm polarizing beam-splitter from Thorlabs, together used to control the power and polarization of laser output. The pulsed operation is made possible by an acousto-optic modulator (AOM) of type AOMO 3200-146 made by Crystal Technology, Inc. The laser light from the polarizing beam-splitter is focused on AOM by a Thorlabs lens (L1) of type AC254-150-AML with a focus distance of 150.0 mm. We use a 573 nm long-pass dichroic mirror (DM) FF573 from Semrock to reflect the excitation light from the laser towards the sample, while transmitting the photoluminescence response of the sample. The dichroic mirror reflects light with a wavelength of less than 573 nm, while light with longer wavelengths can pass through. The excitation light reflected from the dichroic mirror goes through the Olympus MPLANAPO NA0.9 air objective, which focuses it on the sample. From the sample, the photoluminescence is collected by the same objective and passes through the dichroic mirror. Behind the dichroic mirror, a long-pass ET570 filter from chroma is placed and has a cut-off wavelength of 550 nm. It is used to filter out the excitation wavelength. After that, a set of lenses and pinholes are used to block the out-of-focus photoluminescence light allowing us to collect light mainly from the single NV centre. All pinholes used in the setup are Thorlabs P30S pinholes with a diameter of 30 $\pm$ 2 ${\mu}$m. Then the beam is focused by a final lens from Thorlabs (L3), onto an SPCM-AQRH-14 single photon counter from Excelitas Technologies capturing the signal, which is then sent to the control PC.
Piezo stage Physik Instrumente P-545 is used for mapping. The stage holding the sample can be moved with a precision of approximately 10 nm, which makes it possible to build a photoluminescence map in the range of 200 um for each axis.\\
If single defects are visible on the map, the position of the stage can be adjusted to obtain a single defect in focus of the laser light and at the same time in focus of the photodetector.
For microwave application a Ti-Al strip (figure \ref{setup}) is bonded to the chip with the Accelonix TPT HB05 manual wire bonder. The strip is connected to the MW amplifier MiniCircuits ZHL-42+ 1W. The input of the amplifier is connected to the Windfreak SynthHD MW source. Typically 20-25 dBm on a 50 Ohm load is required to observe the resonance on an NV centre 10-30 µm away from the microwave strip. The 50 Ohm terminator is necessary to avoid reflection back into the wire causing interference with the original signal. A small magnetic field, 1-10 G, is enough to split the magnetic resonance of the NV centre - this field is applied with a magnet installed in proximity of the sample (1-5 cm).\\
Titanium-Aluminum (Ti-Al) electrodes and microwave antennas (Fig. \ref{setup}c) are deposited on samples by standard lift-off optical lithography (see Methods). Before being mounted on the sample holder a diamond sample is glued on a chip with metallic strips and SMA connectors attached. The electrodes and the antenna are bonded with the metallic strips allowing to connect voltage source and picoampermeter (Keithley 487 with down to 10 fA sensitivity) to the electrodes and a microwave amplifier and a 50 Ohm load to the antenna.
\section*{Depth scans of a point source}
\begin{figure}[h]
    \centering
    \includegraphics[width=\linewidth]{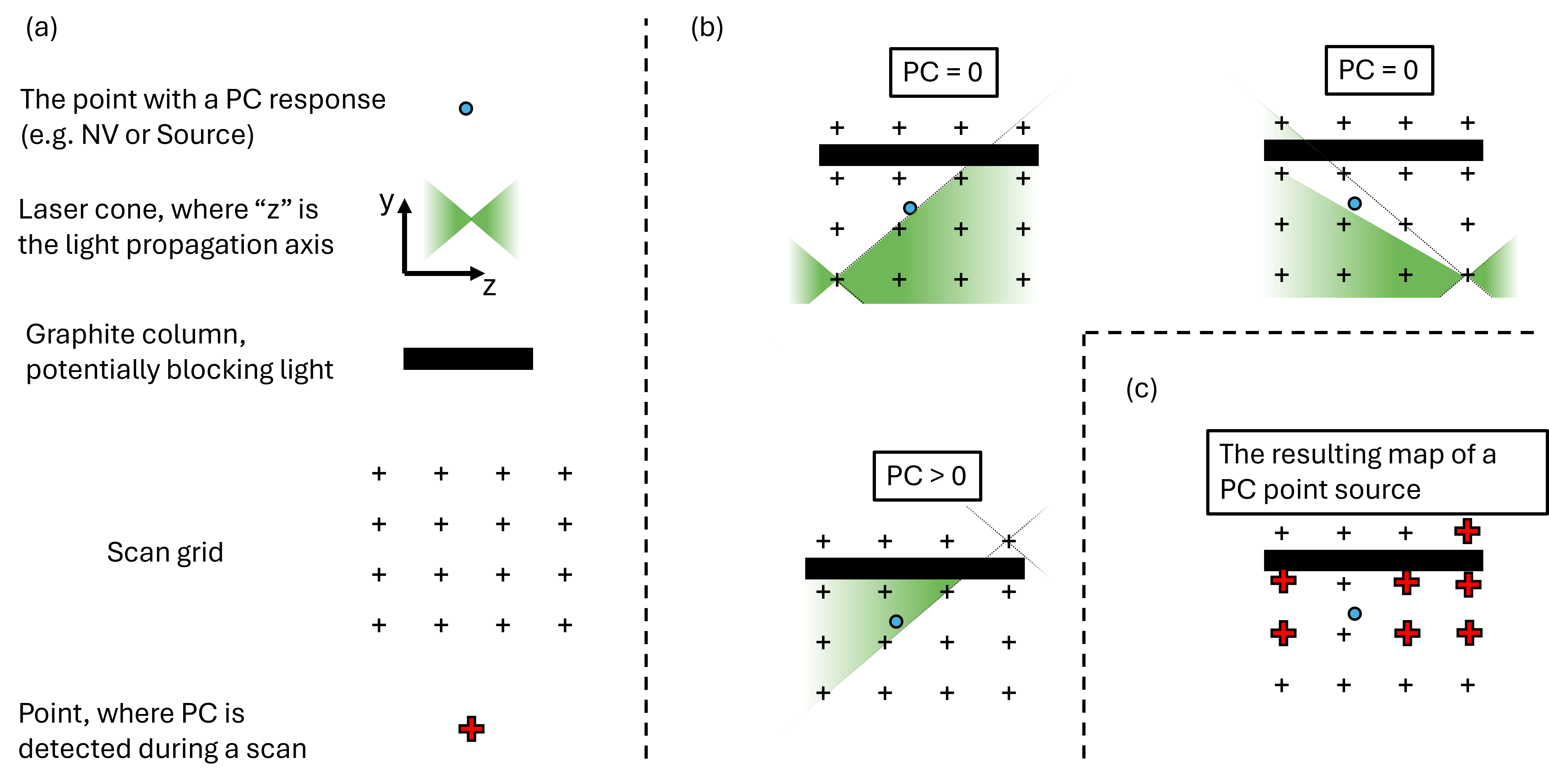}
    \caption{\textbf{(a)} The meaning of the symbols used in the graphical explanation of a depth scan in (b) and (c). \textbf{(b)} PC for various positions of the laser focus with respect to the point source and the graphite column. PC is zero when the point source is not illuminated and larger than zero in any other case. \textbf{(c)} The result of a depth scan around a point source. In depth scans, unlike in surface scans, a point source appears as a cone shape matching the shape of the light beam.}
    \label{yzs}
\end{figure}
Compared to surface scans, the results of depth scans are more difficult to interpret. This is because light propagates along one of the scan axis, and, therefore, moving a source of PC out of the focus point does not guarantee that it is no longer illuminated. This effect in not as apparent on PL scans thanks to the use of a pinhole, which blocks transmission of off-focus PL to a photon detector. A single point in a diamond sample with PC response would appear as a cone shape on a PC depth scan (fig. \ref{yzs}c).
\section*{Distance correction for depth scans}
\begin{figure}[b!]
    \centering
    \includegraphics[width=0.7\linewidth]{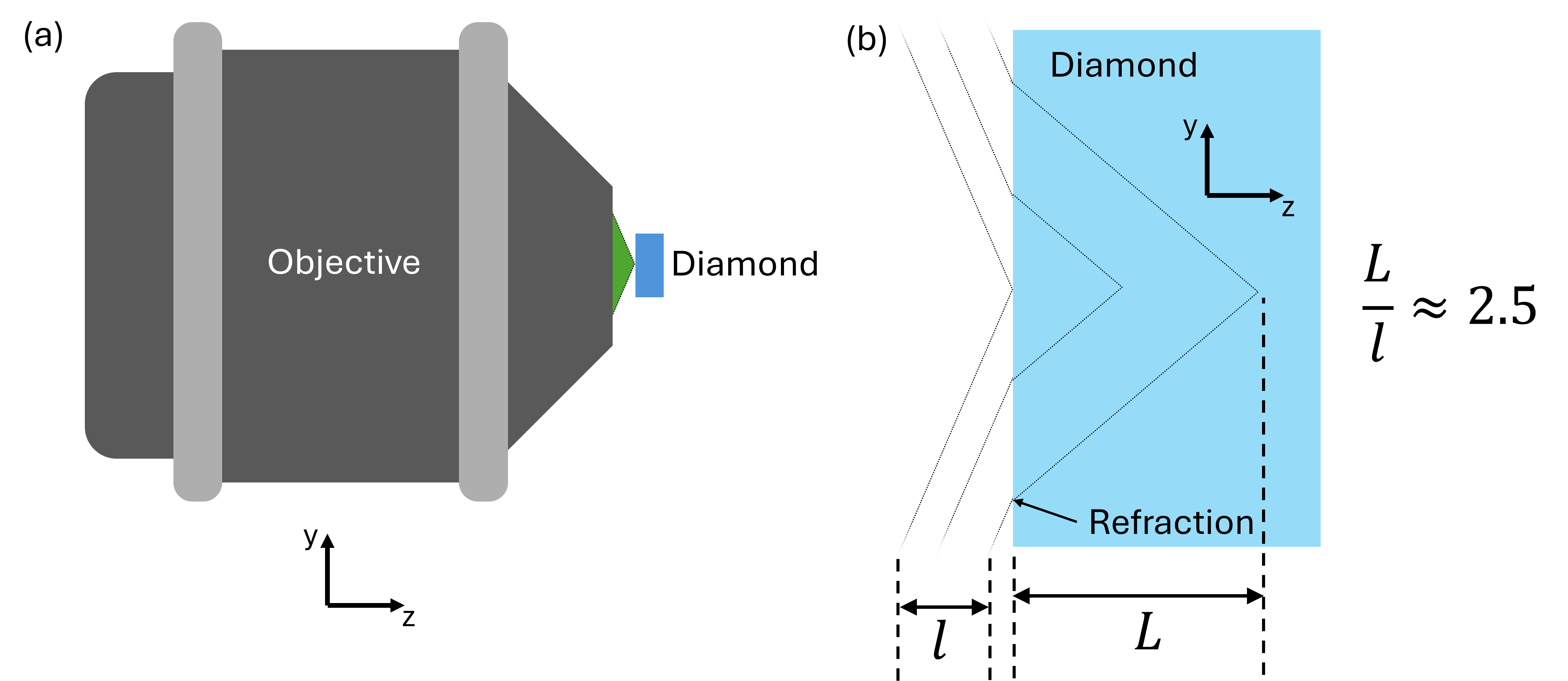}
    \caption{\textbf{(a)} Optical objective focusing laser light on a diamond sample. For a depth scan, the objective moves along the z-axis. \textbf{(b)} The difference between the distance traversed by the objective and the distance traversed by the focus point. Because of the light refraction at the air-diamond interface, the focus point travels significantly faster than the objective. From experimental observations we estimate that the focus point travels approximately 2.5 times faster than the objective.}
    \label{cor}
\end{figure}
During depth scans we record PC as a function of the objective position relative to the sample. However, to calculate the focus point position relative to the sample one needs to account for the refraction of light at the air-diamond interface (fig. \ref{cor}b). From experimental observation we estimate that the correction factor for the distances along z-axis is approximately 2.5. It has to be noted that the focus point becomes smeared as it moves deep into diamond. This happens because laser light hits the surface at a large range of incidence angles all of which get refracted differently.
\section*{Depth scans of Sources}
\begin{figure}[h]
    \centering
    \includegraphics[width=0.7\linewidth]{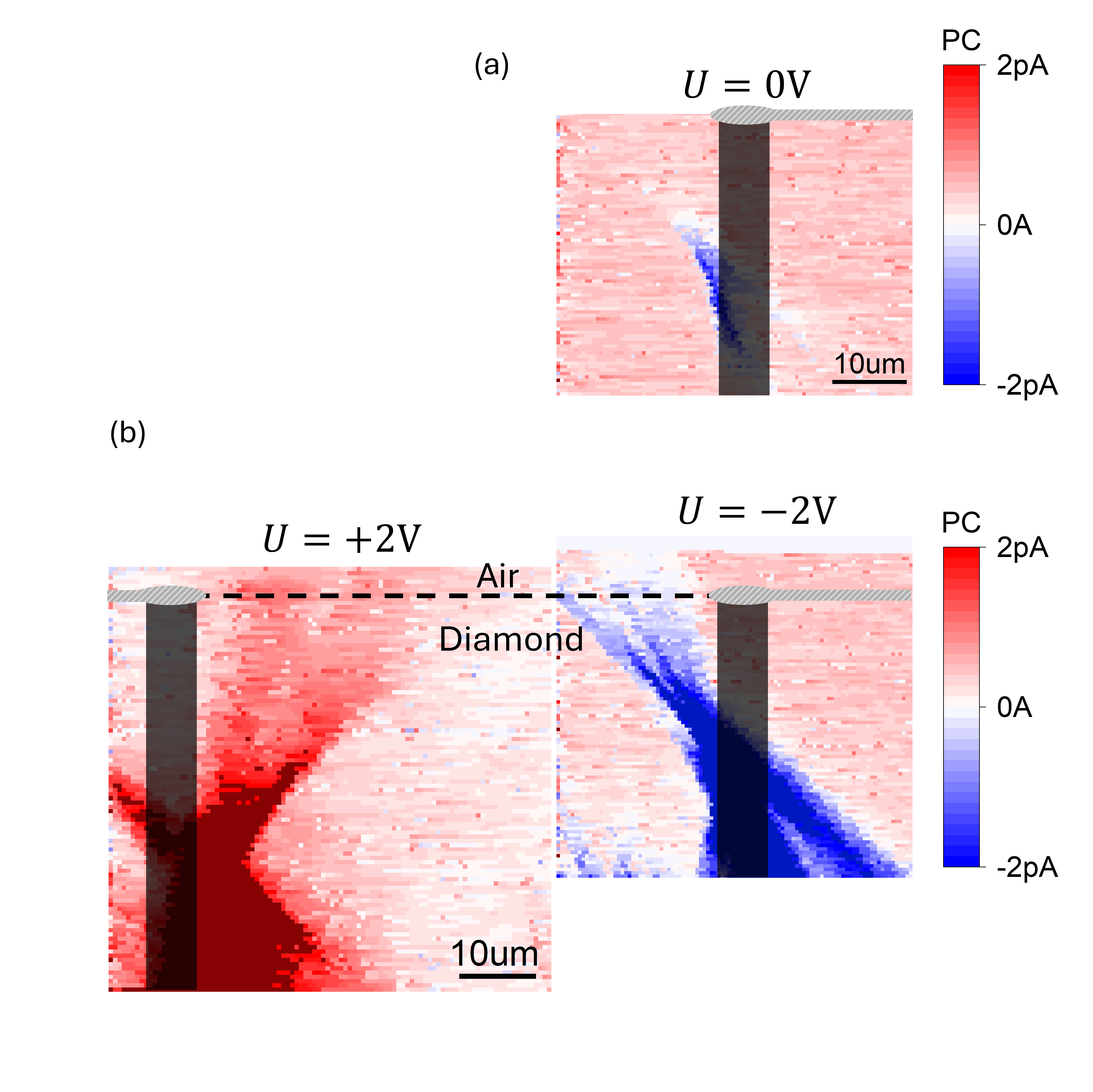}
    \caption{\textbf{(a)} A standard PC depth scan of a Source at 0V. \textbf{(b)} Standard PC depth scans of two Sources on the opposite electrodes. Scans are performed under bias voltage of 2V.}
    \label{injs}
\end{figure}

The two scans in figure \ref{injs}b are made with bias voltage in order to enhance the signal. The shape of the area where PC is observed indicates that there are only two points in the sample, one at each graphite column, which produce PC. The PC that we observe when focusing on the surface of diamond is flowing because laser light illuminates the sample below the focus point as well and at least one of the Sources is illuminated this way. The shape of the photocurrent response indicates that photocurrent is only produced on the condition of either point on the electrode being illuminated.\\
The NV and the source are illuminated by the same laser. NV produces electrons, which populate the Bridge. Holes are being injected at the Source on the condition that the Source is illuminated and the Bridge is not empty.
\section*{IV characteristics of Source PC}
\begin{figure}[h]
    \centering
    \includegraphics[width=\linewidth]{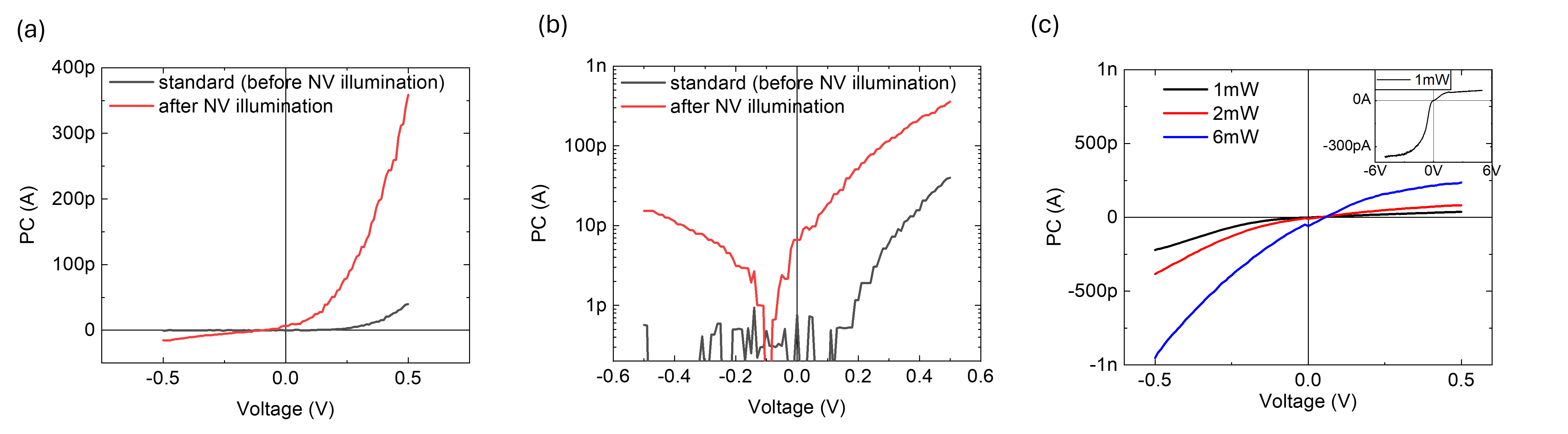}
    \caption{\textbf{(a)} IV curves at a Source before and after illumination of NV centre. Laser power is 3mW. \textbf{(b)} Same as (a) except on a logarithmic scale. \textbf{(c)} IV curves at an NV centre for different laser powers. Inset shows a larger voltage range.}
    \label{ivs}
\end{figure}
We measure IV characteristics, meaning PC dependence on voltage, at a Source (fig. \ref{ivs}a) and at an NV centre (fig. \ref{ivs}b). According to our findings, described in the main text, the PC at the NV centre is actually a combination of PC from the Sources on the diamond-electrode interface. The IV characteristics measured at a Source are similar to an IV characteristic of a diode. It must be noted, however, that there is no detectable current in the dark, which is not characteristic for a diode. It can be concluded from the figure \ref{ivs}a that electrons (either from conduction band or valence band) can enter the electrode (high current at positive bias), but cannot leave (low current at negative bias). Under normal conditions electrode-diamond interface is not conductive in either direction. Laser light creates conditions for conductivity in one direction. Our current hypothesis is that light-assisted charge injection takes place at the interface. We also assume that this process is possible without bias voltage application. The effect of NV centre illumination is not yet completely understood. However, it is clear that NV centre illumination amplifies whatever effect is causing the original PC behaviour
\section*{Discharge measurements}
\begin{figure}[b!]
    \centering
    \includegraphics[width=0.7\linewidth]{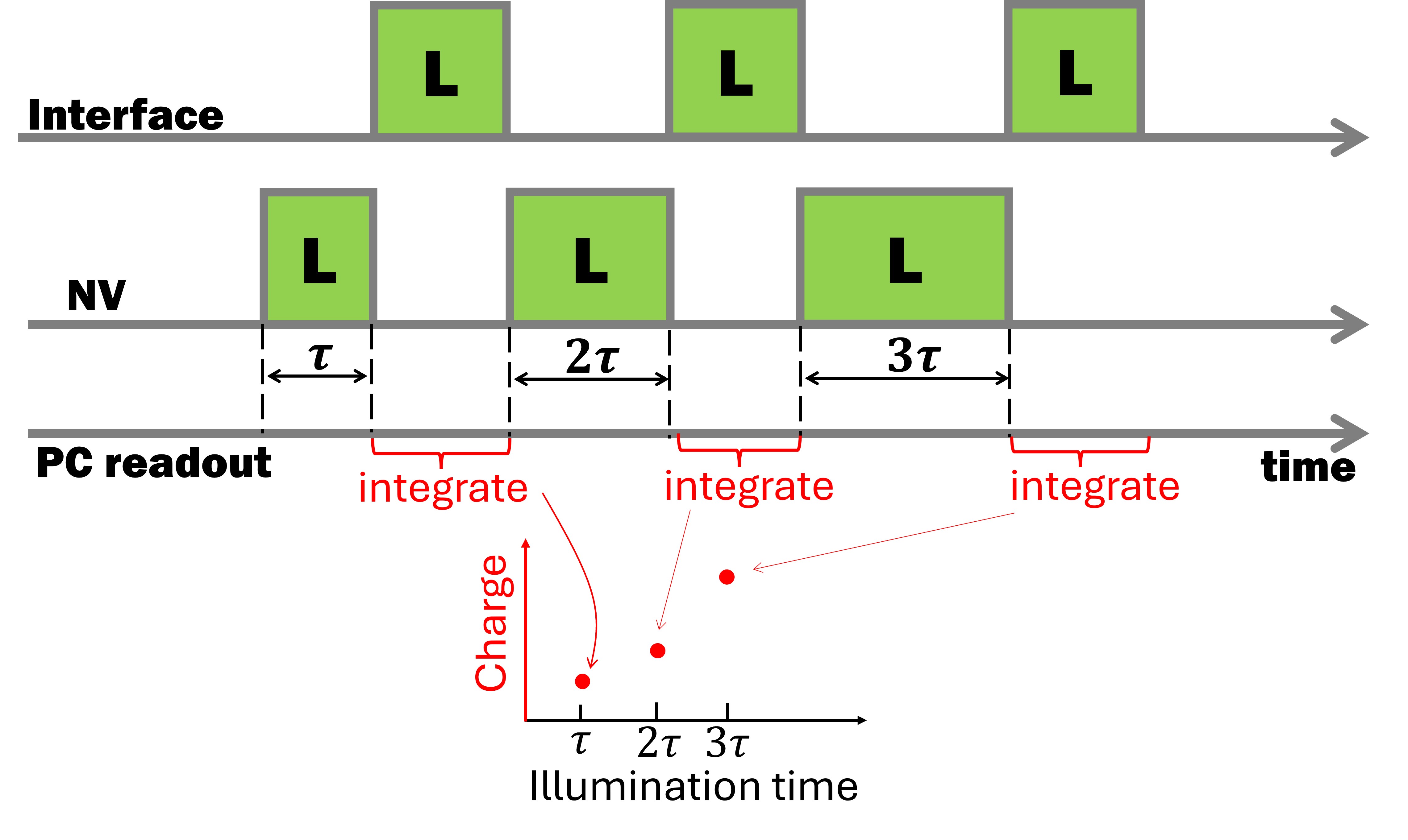}
    \caption{The scheme to measure the charge accumulated at the interface as a function of the NV illumination. NV is illuminated for a varied time before the charge on the electrode is measured by integrating the current from the interface.}
    \label{inch}
\end{figure}
\begin{figure}[b!]
    \centering
    \includegraphics[width=\linewidth]{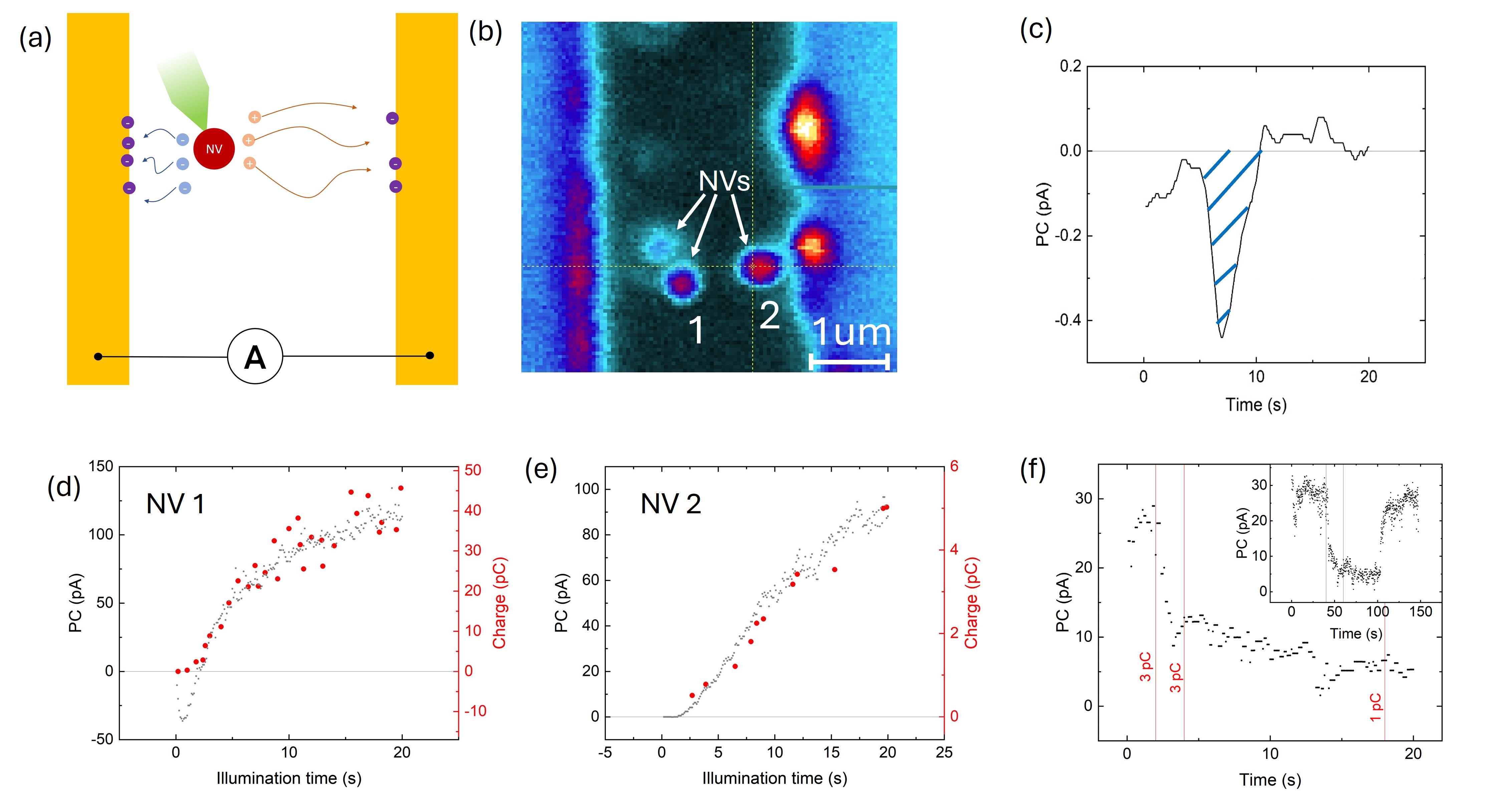}
    \caption{\textbf{(a)} Schematic depiction of the photocurrent and the interface charge. \textbf{(b)} Photoluminescence map of the "in between the electrodes" region of the diamond sample. Locations, which exhibit the discharge effect, also show high photoluminescence under laser excitation. \textbf{(c)} Photocurrent as a function of time during the "discharge" event under illumination. A photocurrent spike is observed at the moment of illumination of the interface of diamond and the electrode. \textbf{(d)} Photocurrent from the NV centres marked in (b) and interface charge at the electrode visible on the left side of (b) in the first 20 seconds after farther electrode discharge for two different NV centres. Apart from the very beginning of NV centre illumination, photocurrent appears to be directly proportional to the interface charge. \textbf{(e)} Same as (e) except on a different NV centre. \textbf{(f)} Recording of photocurrent signal response to on-resonant microwave application. Interface charge is measured directly before, directly after, and significantly after microwave application start (red lines). The initial photocurrent response is not charge induced, the following slow decay, however, most likely is. \textbf{Inset:} The full recording, containing also the response to termination of the on-resonant microwave application. }
    \label{dich}
\end{figure}

A method is developed to observe the amount of an electric charge at the diamond-electrode interface: at the moment of interest we stop illuminating the NV centre, move the focus point (aim) at the interface, turn the laser illumination on again - at this point current starts flowing through the electrodes for about a second, we record this current until it disappears and integrate it (Fig. \ref{inch}). The moment of interest could be e.g. a certain time after resonant microwave radiation is turned on or in the middle of photocurrent initial rise after full electrode discharge. This is an intrusive measurement in the sense that it removes the charge from the interface in order to observe it, therefore experiment pre-conditions had to be reproduced from the start after each such measurement.\\
During our experiments we have noticed that bridge illumination (see main text) is always accompanied by a current spike that lasts for less than a second (Fig. \ref{dich}c). We believe that this is the trace of the charge that has accumulated at the bridge while the NV centre was producing photocurrent. In order to observe this effect once again one has to put a nearby NV centre under laser illumination for some time. This leads us to believe that not all the charge generated by the NV centre is collected by the electrode or recombines - some charge carriers must also populate trap levels on the diamond surface or directly at the interface between titanium and diamond.\\
We measure the correspondence between the recharging process at the farther bridge and the NV photocurrent rise during this recharging. Figures \ref{dich}d and \ref{dich}e show the result of these measurements for two different NV centres. Figure \ref{dich}e shows a measurement for the NV centre farther from the electrode. Both NV centres are marked in figure \ref{dich}b. From these measurements, we conclude that NV photocurrent is mostly proportional to the interface charge.\\
We also investigate the relation between the interface charge and PDMR contrast. We measure the charge before and after microwave application (Fig. \ref{dich}f). The initial photocurrent response to on-resonant microwave application is immediate and dramatic, however the interface charge does not change. For some seconds after the microwave application, the interface charge decreases, which is accompanied by a further drop in photocurrent.
\begin{figure}[h]
    \centering
    \includegraphics[width=0.9\linewidth]{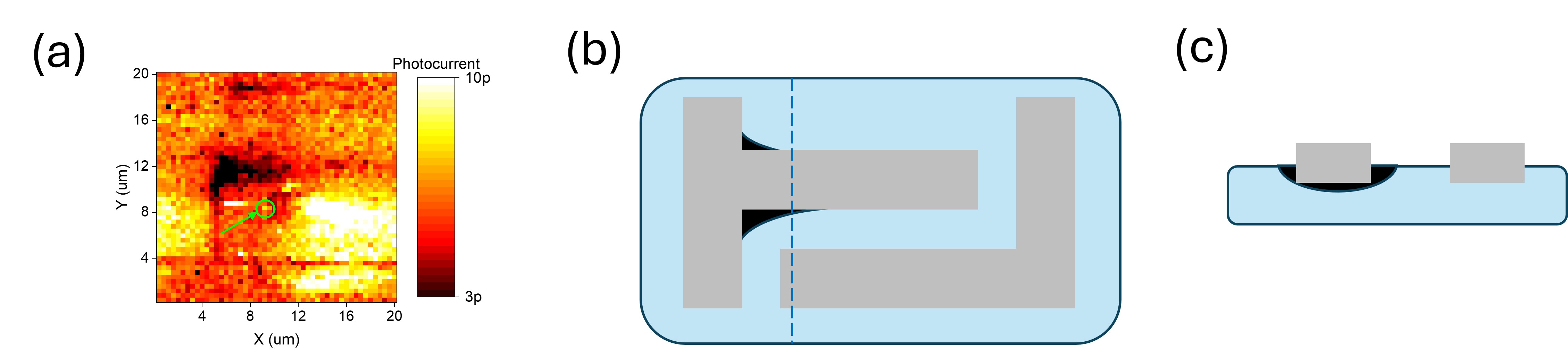}
    \caption{\textbf{(a)} PC reaction map. A second spot is visible at the very top of the scan. \textbf{(b)} The scheme of the spots observed in (a) with respect to the electrodes. \textbf{(c)} Same as (b) except it is side-view at the cut shown in (b) with the dashed line.}
    \label{below}
\end{figure}
\section*{Interface below the electrode}
One of the photocurrent reaction maps shows that the NV PC also reacts to illumination of the other side of the electrode (Fig. \ref{below}a,b). This leads us to believe that the active area of the bridge continues below the electrode (Fig. \ref{below}c), but does not show on the maps because the electrodes are not transparent.
\begin{figure}[h]
    \centering
    \includegraphics[width=0.8\linewidth]{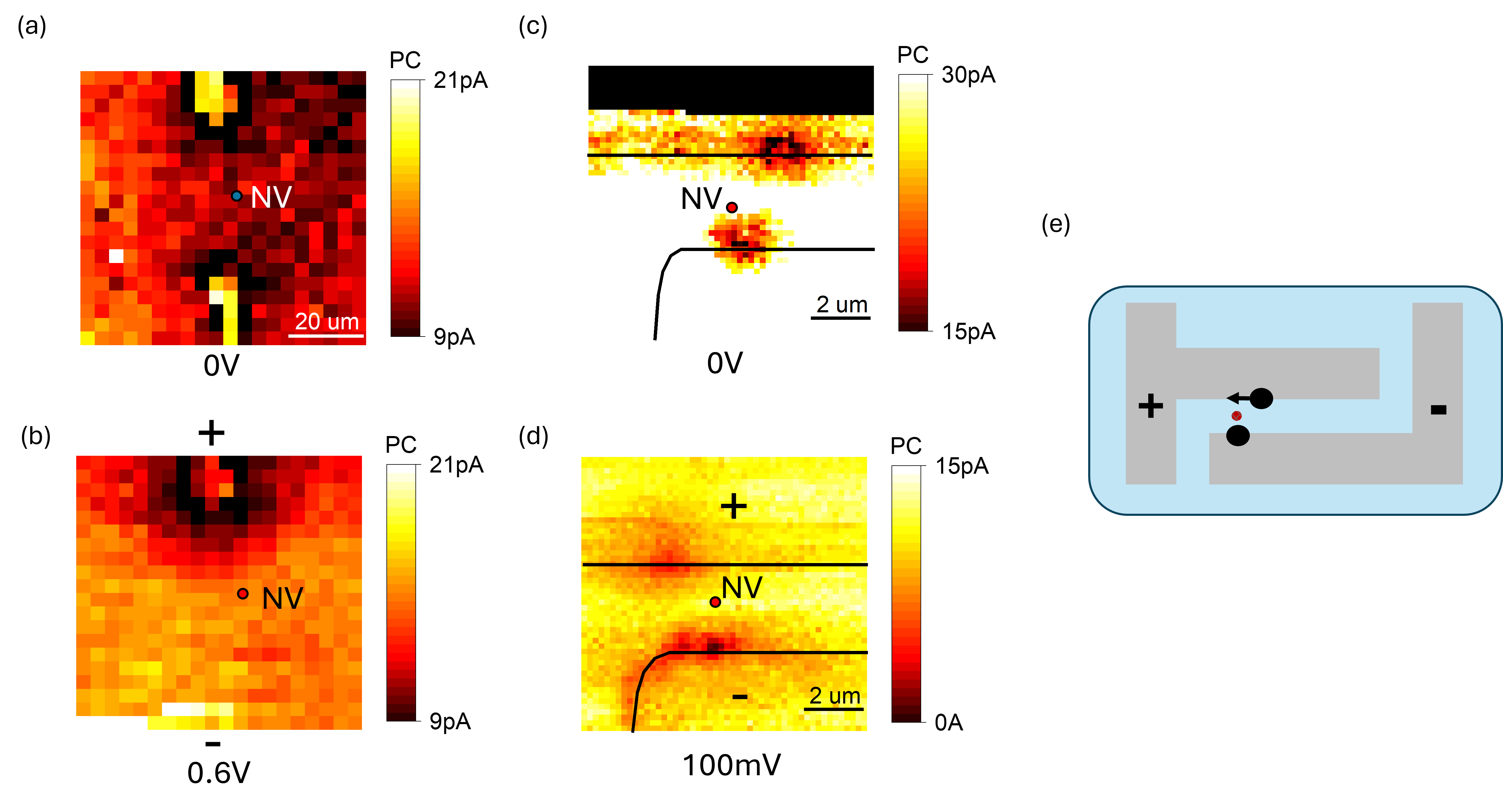}
    \caption{\textbf{(a)} The PC reaction scan of the surface of the Sample G. The PC drop is largest at the interface between diamond and electrode. \textbf{(b)} Same as (a), but under bias voltage. There is no PC drop after illuminating the interface of the cathode, and there is a much stronger drop after illuminating the area around the anode. \textbf{(c)} The PC reaction scan of the surface of the Sample M. There are two locations, one at each electrode, which greatly decrease PC if illuminated. \textbf{(d)} Same as (c), but under bias voltage. One of the spots has shifted to the left. \textbf{(e)} The electrode scheme on Sample M with respect to the spot shift. The upper spot has shifted towards positive bias.}
    \label{br}
\end{figure}
\section*{PC reaction scans under voltage}
PC reaction maps taken under bias voltage show that the bridge position shifts towards the positive bias (Fig. \ref{br}). For this reason, we suspect that the process of building a bridge corresponds to deposition of electrons on the interface energy levels.
\section*{Large scan and laser power dependence}
\begin{figure}[h]
    \centering
    \includegraphics[width=0.7\linewidth]{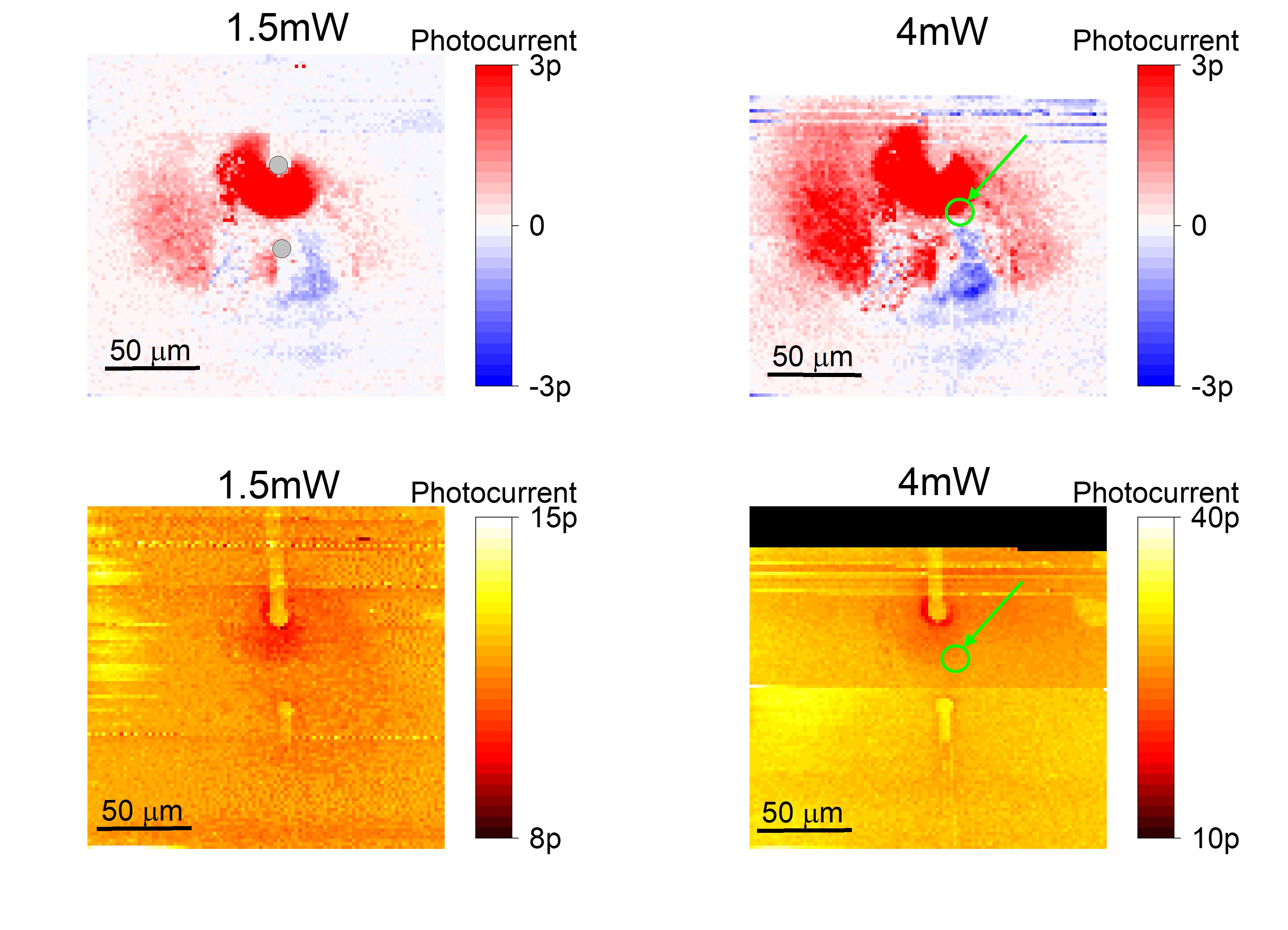}
    \caption{Reset PC scans and PC reaction scans for Sample G at 1.5 and 4 mW laser power. NV is marked with a green arrow. The area of BG PC is significantly larger compared to standard electrodes.}
    \label{gr}
\end{figure}

We measure photocurrent reaction maps for the sample with graphite electrodes. The sample is HPHT bulk diamond with boron concentration up to 100 ppb, irradiated and annealed. Graphite columns are produced with laser treatment. Standard Ti-Al electrodes are deposited on the surface to overlap the graphite columns.\\
We observe mostly the same features as on the other samples, except that the area of BG PC is extended from 20 um to approx. 50 um (Fig. \ref {gr}). A number of smaller differences with standard electrodes can be seen in figure \ref{gr}. However, we believe that a more detailed study is necessary in order to draw any conclusions.\\
We perform the measurements at two laser power values, but observe no qualitative differences.

\section*{Donor-acceptor pair N$_1^+$+B$_1^0$ and PC drop outside of  the interface}
\begin{figure}[h]
    \centering
    \includegraphics[width=0.7\linewidth]{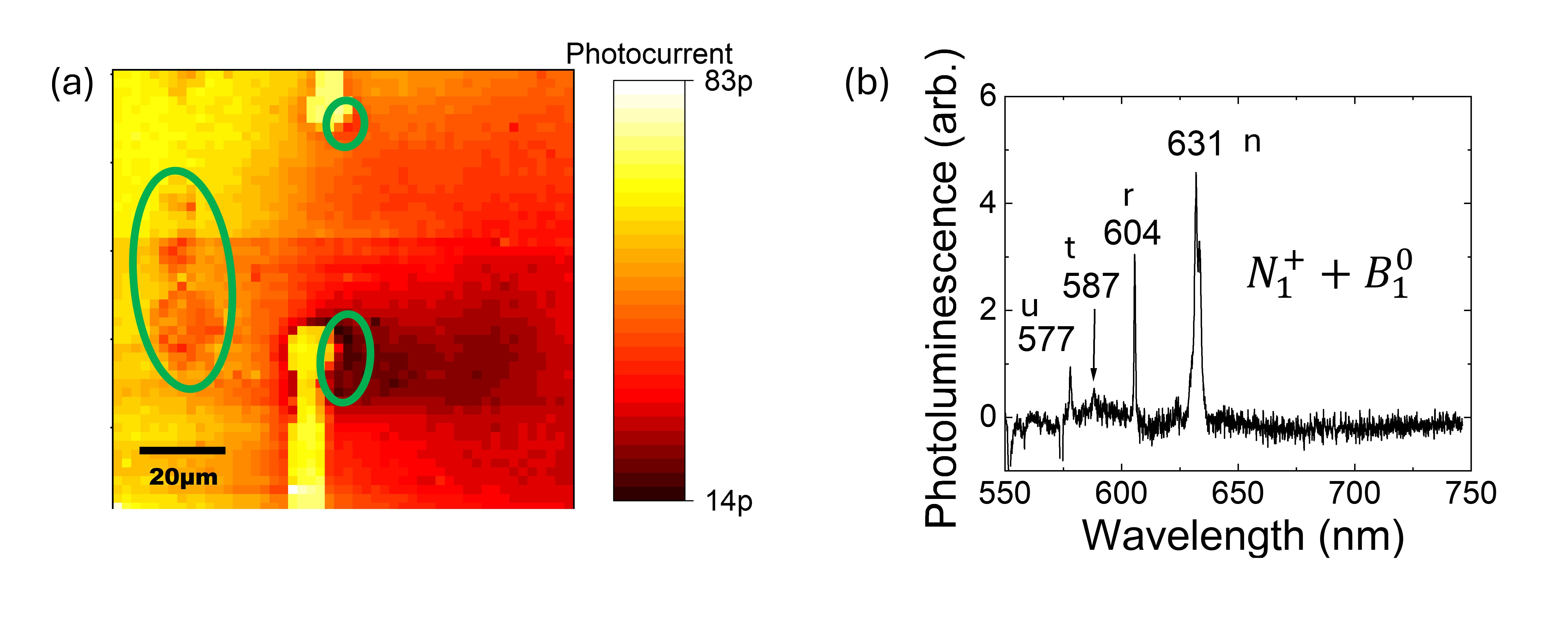}
    \caption{\textbf{(a)} A PC reaction scan on Sample G. At the marked spots a PL spectrum of the N$_1^+$+B$_1^0$ donor-acceptor pair is observed exactly at the points to illumination of which the NV PC reacts the strongest. \textbf{(b)} The PL spectrum of the N$_1^+$+B$_1^0$ donor-acceptor pair observed at the point marked in (a).}
    \label{dap}
\end{figure}
We have discovered that at some locations on the sample, which show a response on the photocurrent reaction maps, (Fig. \ref{dap}a) there is a PL spectrum (Fig. \ref{dap}b) that corresponds to the N$_1^+$+B$_1^0$ donor-acceptor pair (DAP). The relation between the DAP and the Bridge effect requires further investigation.
\section*{The difference between ODMR and PDMR contrast}
Considering that $G\propto IP$ (see main text), we can write the difference in contrast as follows:
\begin{equation}
    \frac{C_\text{PC}}{C_\text{PL}}=\frac{J_\text{off}-J_\text{on}}{J_\text{off}}/\frac{G_\text{off}-G_\text{on}}{G_\text{off}}= \frac{J_\text{off}-J_\text{on}}{G_\text{off}-G_\text{on}}\frac{G_\text{off}}{J_\text{off}}
    \label{c1}
\end{equation}
where $J_\text{off}$, $J_\text{on}$ and $G_\text{off}$, $G_\text{on}$ are the values for the photocurrent and the charge generation rate with and without MW.\\
\begin{figure*}[b!]
    \centering
    \includegraphics[width=0.7\linewidth]{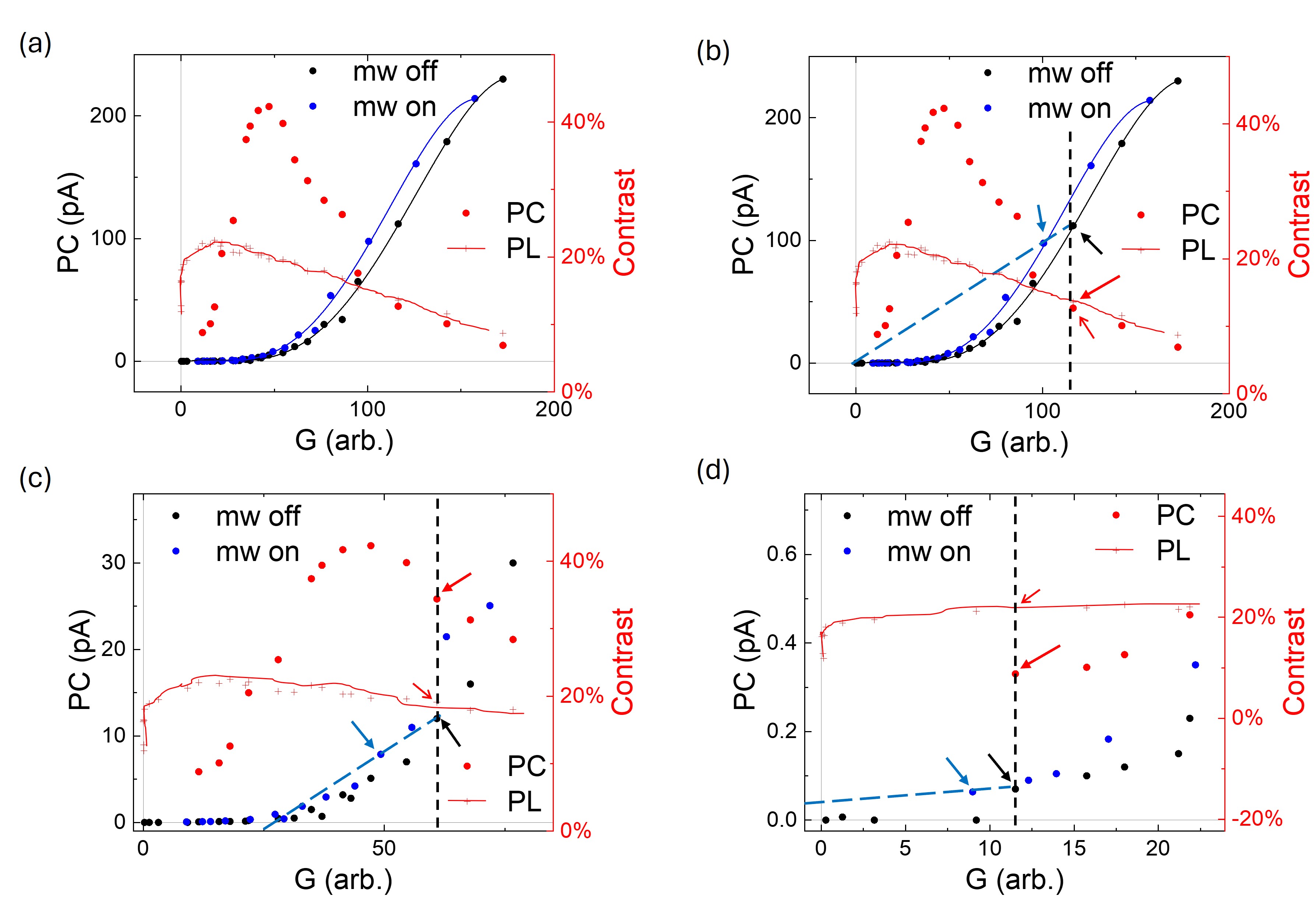}
    \caption{\textbf{(a)} Photocurrent, PC and PL contrast as functions of the charge generation rate. Two important features of the photocurrent are evident - the non-linearity and the dependence on MW. The contrast is plotted with respect to $G_\text{off}$. \textbf{(b)} Relation between the contrast and the features of $J(G)$ at the specific $G$ (vertical dashed line). Considering (\ref{c1}) and (\ref{dbl}), the points $(C_\text{PC},G)$ and $(C_\text{PL},G)$ are marked with the large and small red arrows respectively, the points $(J_\text{off},G_\text{off})$  and $(J_\text{on},G_\text{on})$ are marked with the black and blue arrows respectively, the point ($0$, $B$) is marked with a cross. $B=0$ and $C_\text{PC}=C_\text{PL}$. \textbf{(c)} Same as (b) except $B<0$ and $C_\text{PC}>C_\text{PL}$. \textbf{(d)} Same as (b) except $B>0$ and $C_\text{PC}<C_\text{PL}$.}
    \label{contr}
\end{figure*}
We demonstrate the effects that the features of the $J(G)$ have on the contrast with figure \ref{contr}. The two features that determine the difference between the PC and PL contrast are the non-linearity of $J(G)$ and the dependence of $J(G)$ on MW (Fig. \ref{contr}a). We show this with the dashed blue line in figures \ref{contr}b-d. The equation for the line is
\begin{equation}
    J=\frac{J_\text{off}-J_\text{on}}{G_\text{off}-G_\text{on}} G+B
    \label{dbl}
\end{equation}
where $B$ is a constant that can be found from the condition $J(G_\text{off})=J_\text{off}$. Then
\begin{equation}
    \frac B{J_\text{off}}= 1-\frac{J_\text{off}-J_\text{on}}{G_\text{off}-G_\text{on}}\frac  {G_\text{off}}{J_\text{off}}
\end{equation}
Comparing with (\ref{c1}), we get
\begin{equation}
    \frac{C_\text{PC}}{C_\text{PL}}=1-\frac B{J_\text{off}}
    \label{b}
\end{equation}
From (\ref{b}) and figure (\ref{contr}) we see that the rise of $J(G)$ when MW is on decreases the PDMR contrast, while the non-linearity of $J(G)$ increases it.\\

\section*{Bright spots on the electrodes}
\begin{figure}[h]
    \centering
    \includegraphics[width=\linewidth]{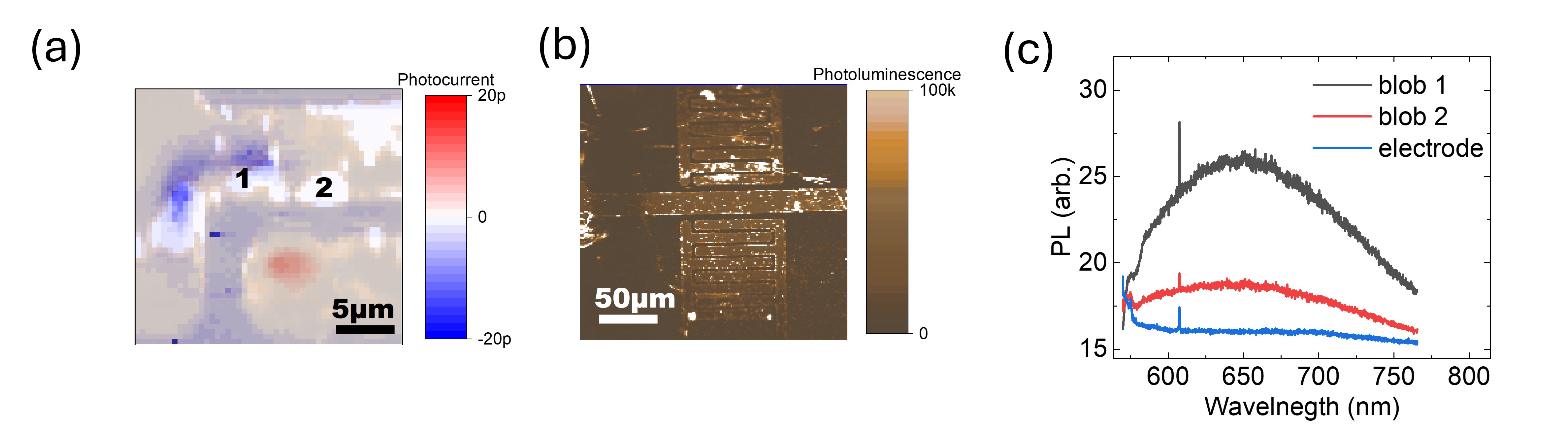}
    \caption{\textbf{(a)} Overlap between a PL map and a PC map. Some of the high PL points coincide with high PC points. \textbf{(b)} PL map of two pairs of interdigitated electrodes with a MW strip in the middle. The top is the working pair of electrodes. The area, on which most of the measurements were conducted is clearly distinguished by the bright spots on the electrodes. \textbf{(c)} PL spectra of the bright spots on the electrode and PL spectrum of the electrode without a spot. The measured points are marked in (a). A weak diamond Raman peak is visible on all spectra at 607 nm.}
    \label{blb}
\end{figure}

We see that prolonged use of the electrodes creates bright spots on them (Fig. \ref{blb}b). Our attempts to correlate the bright spots with the PC from the electrodes unfortunately do not yield conclusive results (Fig. \ref{blb}a). There are some instances of significant overlap between PL and PC, however, for the most part there seems to be no correlation. The spectra of these bright spots (Fig. \ref{blb}c) shed no light on their origin, however the presence of the first diamond Raman peak at 607 nm is an unexpected feature. This implies that aluminium layer is completely or almost completely gone at those points.
\section*{Regions of negative differential resistance}
\begin{figure}[t!]
\centering 
    \includegraphics[width=0.8\linewidth]{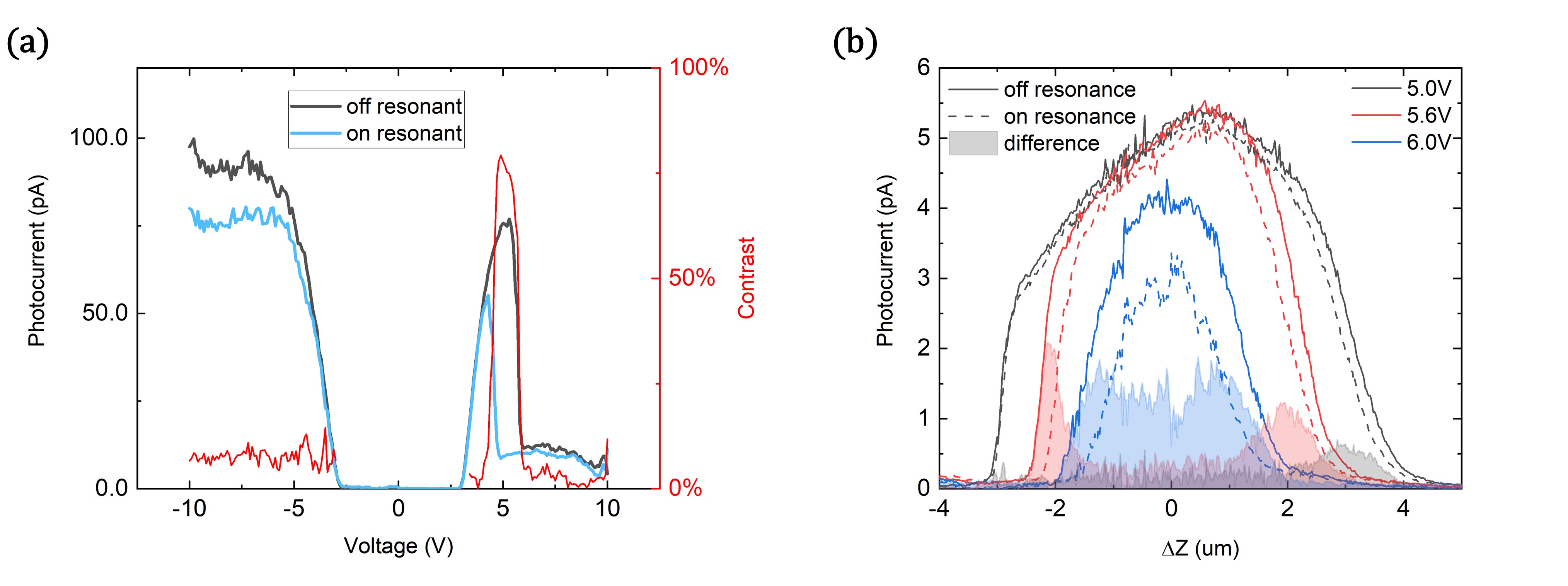}
    \caption{\textbf{(a)} The current-voltage (IV) characteristic of the NV photocurrent. There is a region where photocurrent drops with voltage. This region coincides with the voltage region of highest contrast. \textbf{(b)} Depth scans of the NV photocurrent. NV centre is at the surface of the sample, which corresponds to $\Delta Z = 0$, while $\Delta Z > 0$ corresponds to focusing outside the sample. PDMR contrast increase due to the inversion of the IV characteristic is stronger, or as strong, on the sides, than when focused directly on the NV centre.}
    \label{vz}
\end{figure}
Several times it has been observed that NV centres with high PDMR contrast show anomalous behaviour of their IV characteristics (Fig. \ref{vz}a), the effect lasts for several hours/days. A region becomes observable, where the photocurrent decreases when the voltage increases, the negative differential resistance region, or the inversion region. This region, if it appears, coincides with the highest contrast region of bias voltage. A depth scan of the NV centre vicinity in this voltage region (Fig. \ref{vz}b) reveals that, when the NV centre is on the edge of the focus point, the contrast is the highest. This could be a consequence of the illumination of the surface around NV or a simple dependence on excitation power. Unfortunately, the effect that causes the inversion region is temporary and the reasons for its appearance are unclear, because of this, the PDMR contrast dependence on excitation power in the inversion region is difficult to measure. If it is measured, however, it could explain the behaviour of the PDMR contrast as a function of focus distance presented on Figure \ref{vz}b.
Judging by the photocurrent behaviour presented on the Figure \ref{vz}a and similar measurements, the most likely explanation for the inversion is that under certain conditions a Bridge can be depleted with voltage, resulting in a PC drop.

\section*{Approximations of the PC behaviour}
In this section the following notation is used:
\begin{itemize}
    \item $B$ is the population of the trap levels (which we will call Bridge population)
    \item $G$ is the charge carrier generation rate from the NV centre
    \item $P_\text{B}$ is the power of Bridge illumination
    \item $n_\text{rec}$ is the population of the free charge carriers which can depopulate the trap via recombination
    \item $J$ is Source PC (which we consider to be the same as the total PC)
    \item $ P_\text{S}$ is the power of Source illumination
    \item $J_\text{S}(U)$ is the current-voltage characteristic of a Source when $B=0$
    \item $a$, $b$, $c$, $d$ are free parameters
\end{itemize}
All the parameters are unitless.\\
We consider three mechanisms to influence the population of the trap levels of the Bridge: 
\begin{enumerate}
    \item A free charge carrier from NV centre is captured by a trap, ionizing the trap. The contribution of this mechanism can be expressed as a product of the number of free charge carriers and empty trap levels - $G^a\left(B_\text{max} - B\right)$. Here, the capture cross-section coefficient is omitted. We expect that $a=1$ since a single charge carrier is captured by a trap independently of other charge carriers (superposition principle). Yet, we keep $a$ is a free parameter to see if the experimental data would confirm that $a=1$.
    \item A photon is captured by a trap, deionizing the trap. The contribution of this mechanism can be expressed as a product of the number of photons and full trap levels - $P_\text B^bB$. Here, the capture cross-section coefficient is omitted. We expect that $b=1$ or $b=2$, depending on the number of photons required to depopulate the trap.
    \item A free charge carrier of the opposite polarity is captured by a trap, deionizing the trap. The contribution of this mechanism can be expressed as a product of the rate of creation of free charge carriers of the opposite polarity and full trap levels - $n_\text{rec}B$. Here, the capture cross-section coefficient is omitted.
\end{enumerate}
We do not consider recombination of the free charge carriers with each other, as we expect its contribution to be negligible compared to the other mechanisms due to large diamond band gap of 5.5 eV. The three mechanisms we consider result in the following rate equation for the Bridge population:
\begin{equation}
    \dot B= G^a\left(B_\text{max} - B\right) - P_\text B^bB - n_\text{rec}B
    \label{b0}
\end{equation}
We have observed that populating the Bridge increases PC by approximately the same factor for any voltage (fig. \ref{ivs}b). For this reason, we expect the dependence between PC and $B$ to be a follows:
\begin{figure}[h]
    \centering
    \includegraphics[width=0.6\linewidth]{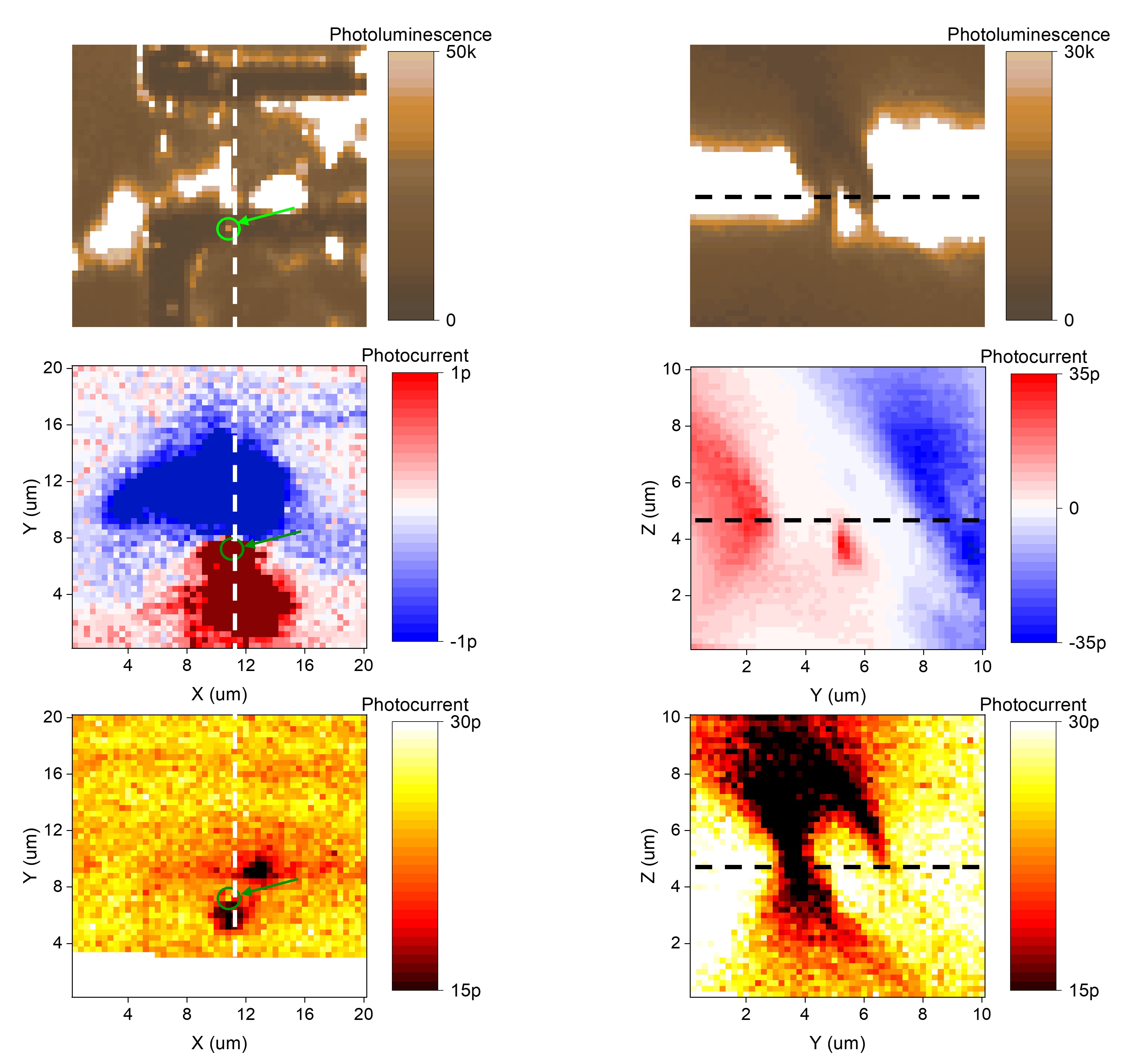}
    \caption{PL, PC, and PC reaction maps from top to bottom. Surface and depth scan from left to right. NV is marked with a green arrow. The cut of the depth scan is marked on the surface scan with the dashed line and vice-versa. 
    The conical shape that we see on PC and PC reaction maps corresponds to the beam profile and indicates that the the observed effects are surface effects.}
    \label{xz}
\end{figure}
\begin{equation}
    J=(B^c+1)P_\text{S}^d J_\text{S}(U)
    \label{j1}
\end{equation}
where coefficient $c$ represents an ambiguity in the relation between PC and Bridge population. In principle, the correct analytical representation of this relation could be much more complex, however, we do not aim here to estimate it beyond first-order approximation. Here, the capture cross-section coefficient is omitted. We expect that $d=1$ or $d=2$, depending on the number of photons required to depopulate the trap.
\section*{2D depth scans}
We try to observe how the Bridge effect depends on the depth of the laser focus. The 2D depth scans that we take (Fig. \ref{xz}), reveal the shape of the signal that corresponds more or less to the beam shape, form which we conclude that this is a surface effect.

\section*{Scans and contrast}
\begin{figure}[h]
    \centering
    \includegraphics[width=\linewidth]{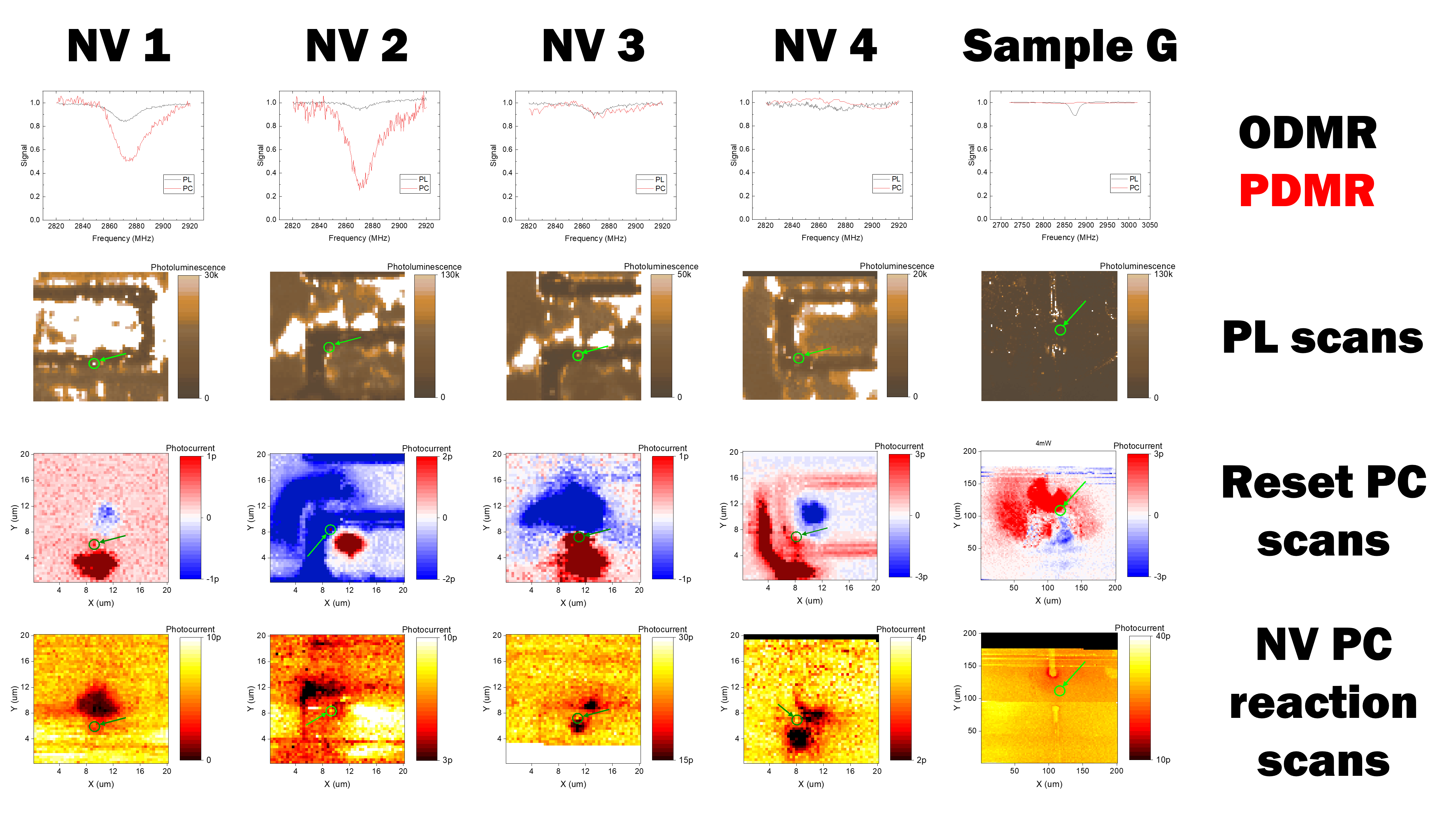}
    \caption{PL, Reset PC, and PC reaction scans and the corresponding ODMR and PDMR contrasts. PDMR contrast is low when two dark spots are observed and high when only one dark spot is observed.}
    \label{scans}
\end{figure}

Figure \ref{scans} shows that NV centres which are close to the boundary between electrode and diamond exhibit high PDMR contrast and for those NVs only one Bridge is visible on the NV PC reaction scans. This does not mean that the second Bridge does not exist, it is more likely that it exists, yet is constantly illuminated due to proximity to the NV centre. Constantly illuminated Bridge would not show on NV PC reaction scans.

\begin{figure}
    \centering
    \includegraphics[width=0.6\linewidth]{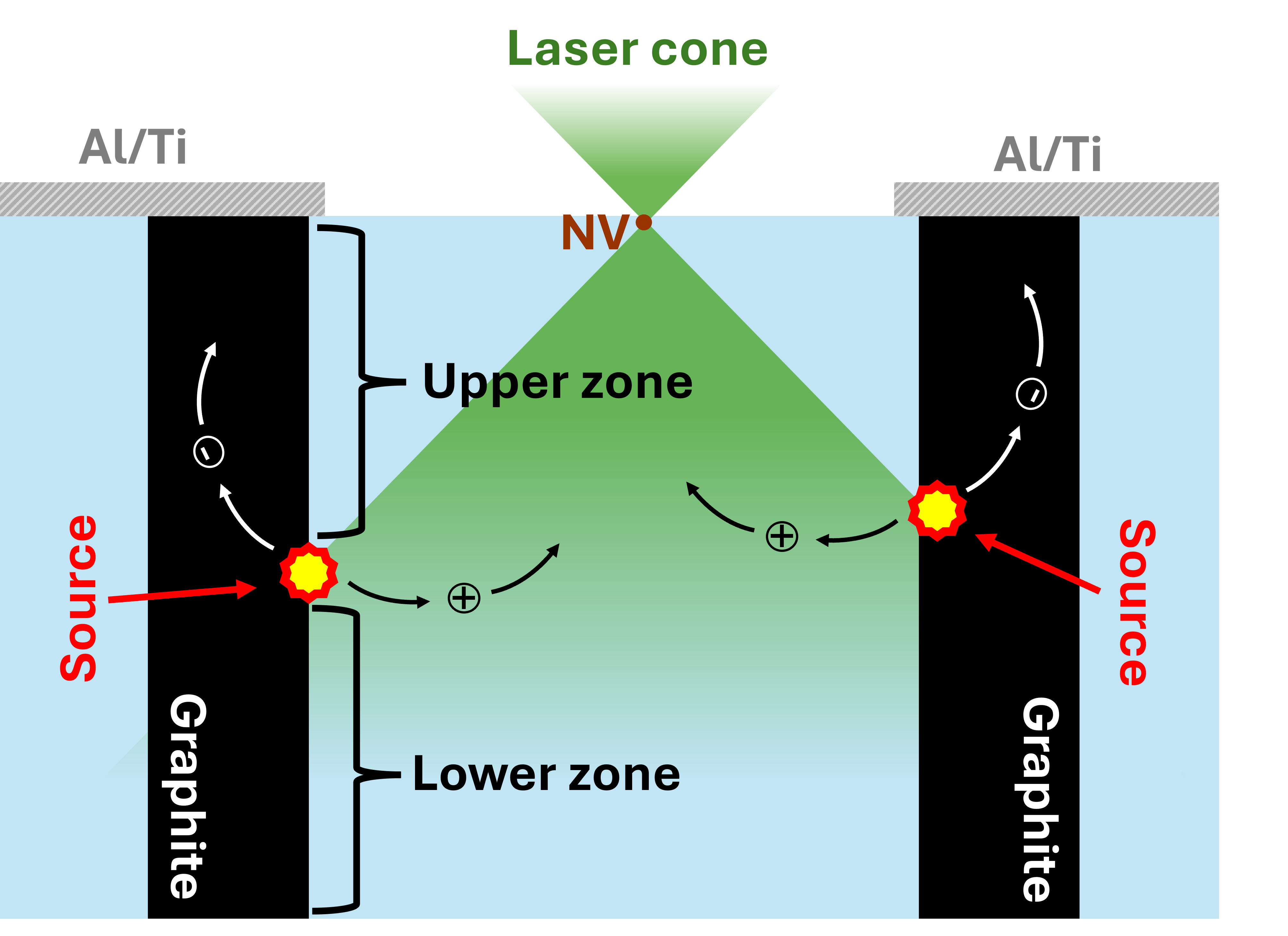}
    \caption{Schematic of Sample G under illumination. Laser illuminates the NV centre together with the graphite columns. Sources are found at the closest to the top surface illuminated points on the graphite columns.}
    \label{sources}
\end{figure}
\section*{On the origin of the localized nature of Sources}
We find that Sources for Sample G are localized to points on the graphite columns. We observe the following features of these Sources:
\begin{enumerate}
    \item Sources for Sample G coincide with the illumination pattern for the case when a particular NV centre is in focus (fig. \ref{sources}). The NV in question was illuminated extensively during the initial sample characterisation.
    \item Sources are the points on the graphite columns closest to sample surface, considering only illuminated points.
\end{enumerate}
From these observations, we make the assumption that continuous transfer of charge through the graphite-diamond interface creates favourable conditions for further transfer charge through the interface, creating the cascade effect. In this picture, the Sources have appeared on the electrodes during the initial illumination of the NV centre. The points in the upper zone (fig. \ref{sources}) could not transfer charge during the initial NV illumination since these points were not illuminated. The points in the lower zone (fig. \ref{sources}) could transfer charge, however this would be a longer path for a charge carrier travelling in and out of sample, and therefore unfavourable. Prolonged illumination of the NV centre has caused the points of the highest charge transfer rate to become favourable for charge transfer in all of the following measurements.
